\documentclass[twocolumn,floats,showpacs,prd,amsmath,amssymb,floatfix,nofootinbib,balancelastpage]{revtex4}
\usepackage{psfig}

\newcommand{\ApJL}{Astrophys. J. Lett.}

\newcommand{\ApJ}{Astrophys. J.}
\newcommand{\PRL}{Phys. Rev. Lett.}
\newcommand{\PRD}{Phys. Rev. D}
\newcommand{\PRA}{Phys. Rev. A}
\newcommand{\PR}{Phys. Rev.}
\newcommand{\MNRAS}{Mon. Not. Roy. Astron. Soc.}
\newcommand{\AsAs}{Astron. Astrophys.}

\newcommand{\vhat}[1]{\hat{{\bf #1}}}
\newcommand{\vect}[1]{{\bf #1}}

\newcommand{\pphi}{\partial_{\varphi}}

\newcommand{\<}[1]{\langle {#1} \rangle}

\newcommand{\conv}{{ \bf \star}}
\newcommand{\convs}{{ \bf \star_{s}}}
\newcommand{\convc}{{ \bf \star_{c}}}

                              \newlength{\strikewidth}
                              \newlength{\strikelength}
\begin{document}

\title{Spatial variation of the fine-structure parameter and the \\ cosmic
microwave background}
\author{Kris Sigurdson}
\email{ksigurds@tapir.caltech.edu}
\affiliation{California Institute of Technology, Mail Code 130-33, Pasadena, CA
91125}
\author{Andriy Kurylov}
\email{kurilov@krl.caltech.edu}
\affiliation{California Institute of Technology, Mail Code 130-33, Pasadena, CA
91125}
\author{Marc Kamionkowski}
\email{kamion@tapir.caltech.edu}
\affiliation{California Institute of Technology, Mail Code 130-33, Pasadena, CA
91125}


\begin{abstract}

We study the effects on cosmic microwave background (CMB) temperature
and polarization anisotropies of spatial fluctuations of the
fine-structure parameter between causally disconnected regions
of the Universe at the time of recombination.  Analogous to weak
gravitational lensing, in addition to modifying the mean power
spectra and inducing a curl component (B mode) to the polarization, spatial
fluctuations of the fine-structure parameter induce higher-order
(non-Gaussian) temperature and polarization correlations in the
CMB.  We calculate these effects for the general case of
arbitrary correlation between temperature fluctuations and
fine-structure parameter fluctuations, and show the results for
a model where these two types of fluctuations are uncorrelated.
The formalism we
present here may also be applied to other
modifications of recombination physics that do not significantly
alter the evolution of the dominant density perturbations.  We
discuss the constraints on the effective Lagrangian for variable
fine-structure parameter necessary to realize this scenario.
\end{abstract}



\maketitle

\section{Introduction}
The possibility that the fine-structure parameter $\alpha$ may
vary in time has long been entertained
\cite{Jordan37,Teller48,Dicke59,Stanyukovich63, Dicke65, Gamow67a,Forgacs79,Beken82}, and
has received renewed interest with recent evidence
{}from quasar spectra that may support a variation of
less than one part in $10^4$ over a time scale of $\sim
10$ Gyr \cite{Webb02,Bahcall03}.  Although the results may still be
controversial, the observational work has inspired
theoretical work that investigates models with
variable $\alpha$
\cite{Uzan03,Maretal03,Battye01,Sandvik02,Youm02,Barrow02d,Kostelecky02,Huang03}, as well as
other work that investigates
possible connections with dark energy and new long-range forces
\cite{Dvali02,Bek02,Chiba02}.  It has also stimulated a
more careful re-investigation of the constraints placed on variable
$\alpha$ from big-bang nucleosynthesis \cite{Nollett02}.

Of course, if a relativistic theory allows for temporal
variation of $\alpha$ then it must also allow for spatial variations of
$\alpha$ between regions not in causal contact.  In
this paper we study cosmological probes of spatial variations in
$\alpha$, focusing in particular on the cosmic microwave
background (CMB), which is rapidly becoming an increasingly
precise probe of cosmological models
\cite{Junetal96,deBernardis00,Spergel03}, as
well as the physics that underlies them
\cite{Kamion99}.  As we will see, spatial variation
in $\alpha$ induces a spatially-varying power spectrum.  This induces ``non-Gaussian'' signatures in the CMB, in the form
of locally anisotropic correlation functions, that cannot be
described by the power spectrum alone (although strictly
speaking the joint probability distribution of temperatures at
$n$ points remains a multivariate Gaussian). Quite interestingly, these effects are analogous to those of weak gravitational lensing
(``cosmic shear'') on the CMB, but differ in detail.  So, for
example, spatial variation of $\alpha$ can alter slightly the
CMB power spectrum, induce a curl component (B mode) in the
polarization, and induce higher-order temperature and
polarization correlations in the CMB.  Effects like those we
investigate here may also arise if there are other spatial
variations in recombination physics that do not involve significant
density/pressure perturbations.  Our calculations are thus
illustrative theoretically, apart from the specific application
on which we focus.

Below, we first review previous work \cite{Hann99,Kap99} that shows how the
recombination history depends on a homogeneous shift in $\alpha$
and then discuss how it affects the CMB power spectrum.  We then
show how spatial variations of $\alpha$, in which the mean value
of $\alpha$ is unaltered, affect the CMB temperature and
polarization power spectra, and in so doing show that a curl
component is induced in the CMB polarization.  We then calculate
the CMB bispectrum and trispectrum induced by spatial $\alpha$ variation.
Throughout, we compare with the analogous calculations for weak
lensing, and show how the effects of weak lensing and spatial
$\alpha$ variations differ.  We then discuss the properties of
and constraints on a toy field-theory model for spatial $\alpha$
variation that produces the CMB
effects we investigate here, without inducing significant
density perturbations.

Before proceeding further, we clarify that here we investigate
spatial variations in the fine-structure parameter $\alpha=e^2
/\hbar c$ that arise {\it only} from spatial variations in the
electromagnetic gauge coupling $e$; we do not tinker with relativity
nor quantum mechanics.
\section{Recombination and {\large $\alpha$}}
\label{sec:Recomb}

Recombination depends on the value of $\alpha$ because the visibility function,
the probability distribution of when a photon last scattered, is
dependent on $\alpha$.  The visibility function is defined as
\begin{equation}
g(t)=e^{-\tau}\frac{d \tau}{dt} \, ,
\end{equation}
where
\begin{equation}
\frac{d \tau}{dt}=x_{e}n_{p}c\sigma_{T} \, ,
\end{equation}
is the differential optical depth of photons due to Thomson scattering.
Here,
\begin{equation}
\sigma_{T}=\frac{8\pi \hbar^2 \alpha^2}{3m_{e}^2 c^2} \, ,
\end{equation}
is the Thomson cross section, $n_{p}$ is the total number density of protons
(both free and bound), and $x_{e}$ is the fraction of free electrons. The
strongest effect of variations of $\alpha$ on this quantity occur due to the
alteration of the
ionization history $x_{e}(t)$.

The recombination of hydrogen cannot proceed through direct recombination to
the ground state
because the emitted photon will immediately ionize a neighboring atom with high
probability.  Instead, the ionized fraction decreases mainly through the
two-photon process $2s \rightarrow 1s$, or via the cosmological redshifting of
$2p \rightarrow 1s$ Lyman-$\alpha$ photons out of the Lyman-$\alpha$ line.
These processes are described by a single differential equation
\cite{Peebles68},
\begin{equation}
\frac{dx_{e}}{dt}={\cal C} \left[ \beta
{e}^{-\frac{B_{1}-B_{2}}{k_{B}T}}(1-x_{e})- {\cal R} n_{p}x_e^2 \right] \, ,
\label{eqn:recomb}
\end{equation}
where $\beta$ is the ionization coefficient, ${\cal R}$ is the recombination
coefficient, ${\cal C}$ is the Peebles efficiency factor (discussed below), and
\begin{equation}
B_{n}=\frac{m_{e}c^2 \alpha^2}{2n^2} \, ,
\label{eqn:binding}
\end{equation}
is the binding energy of the level with principle quantum number $n$.

Through the Einstein relations we relate $\beta$ to ${\cal R}$ through
\begin{equation}
\beta=\left(\frac{m_{e}k_{B}T}{2\pi
\hbar^2}\right)^{\frac{3}{2}}{e}^{-\frac{B_{2}}{k_{B}T}}{\cal R} \, .
\end{equation}
The recombination coefficient can be written
\begin{equation}
{\cal R}=\sum_{n=2}^{\infty} \sum_{l=0}^{n-1} \alpha_{nl}w_{n} \, ,
\end{equation}
where
\begin{equation}
\alpha_{nl}=\frac{8\pi(2l+1)}{(2\pi m_{e}
k_{B}T)^{\frac{3}{2}}c^2}{e}^{\frac{B_{n}}{k_{B}T}}\int_{B_{n}}^{\infty}
d(h\nu) \sigma^{bf}_{nl} \frac{(h\nu)^2}{{e}^{\frac{h\nu}{k_{B}T}}-1} \, ,
\end{equation}
is the rate at which atoms recombine to the $n,l$ energy level and
$w_{n}$ is the efficiency for an $n$ level to survive in a plasma
\cite{Boschan98}.  The details of $w_{n}$ are not important for the present
discussion, other than to note that at the densities of interest it is unity
for $ n < n_{max} \sim 500$.  Since the dominant contributions to ${\cal R}$
come from $n \lesssim 50$, ${\cal R}$ is insensitive to the weak $\alpha$
dependence of $n_{max}$.  Above, $\alpha_{nl}$ is written in terms of the
ionization cross section $\sigma^{bf}_{nl}$, which can be expressed in the form
\cite{KarzasRybicki}
\begin{equation}
\sigma^{bf}_{nl}=\alpha^{-1}f_{n}\left(\frac{h\nu}{B_{1}}\right) \, .
\end{equation}

Thus, we can write
\begin{align}
\alpha_{nl}=\frac{8\pi(2l+1)}{\alpha c^2}&\left(\frac{k_{B}T}{2\pi m_{e}}\right)^{\frac{3}{2}}{e}^{\xi
/n^{2}} \nonumber \\
 &\times \int_{\xi /n^{2}}^{\infty} dy
f_{n}\left(\frac{y}{\xi}\right)\frac{y^2}{{e}^{y}-1}, 
\end{align}
where
\begin{equation}
\xi =\frac{B_{1}}{k_{B}T}=\frac{m_e c^2 \alpha^2}{2k_{B}T} \, ,
\end{equation}
and it immediately follows that the $\alpha$ dependence of ${\cal R}$ is of the
form,
\begin{equation}
{\cal R}=\alpha^{-1}T^{\frac{3}{2}}F(\xi)=\alpha^2G(\xi)\, .
\label{eqn:alphadep}
\end{equation}

As shown in Ref.~{\cite{Ma95}}, for the temperatures of interest,
\begin{equation}
{\cal R}=\frac{64}{3}\frac{\hbar^2 \alpha^2}{m_{e}^2 c}\sqrt{\frac{\pi
B_{1}}{3k_{B}T}}\phi_2 \, ,
\end{equation}
where
\begin{equation}
\phi_2 \simeq \frac{13\sqrt{3}}{16\pi}{\rm ln}\left(\frac{B_{1}}{k_{B}T}\right)
\, .
\end{equation}
Given the scaling in Eq.~(\ref{eqn:alphadep}), we can then read off the
$\alpha$ dependence. Explicitly, the recombination
coefficient is
\begin{equation}
{\cal R}=\frac{52}{3}\frac{\hbar^2 \alpha^3}{\sqrt{2 \pi m_{e}^3 k_{B}T}}{\rm
ln}\left(\frac{m_{e}c^2 \alpha^2}{2k_{B}T}\right) \, .
\end{equation}

The rate of recombination is inhibited by ionizing photons which can
disrupt atoms in the $n=2$ state before they can decay to the ground state.
The efficiency of recombination from the $n=2$ state is described by the
Peebles efficiency factor,
\begin{equation}
{\cal C}=\frac{\Lambda_{H} + \Lambda_{2s \rightarrow 1s}}{\Lambda_{H} +
\Lambda_{2s \rightarrow 1s} + \beta} \, ,
\label{eqn:peebles}
\end{equation}
which is just the ratio of the recombination rates to the sum of the
recombination and ionization rates from the $n=2$ level.

In Eq.~(\ref{eqn:peebles}),
\begin{equation}
\Lambda_{H}=\frac{8\pi H}{(\lambda_{2p \rightarrow 1s})^3 n_{1s}} \, ,
\end{equation}
is the rate at which a recombination is successful because the emitted
Lyman-$\alpha$ photon is redshifted out of the Lyman-$\alpha$ line before
ionizing a hydrogen atom.  Here, $H=({1}/{a})({da}/{dt})$ is the Hubble
expansion rate,
$n_{1s} \simeq (1-x_{e})n_{p}$ is the number density of atoms in the $1s$ state
(almost all hydrogen atoms are in the $1s$ state), and
\begin{equation}
\lambda_{2p \rightarrow 1s}=\frac{16 \pi \hbar}{3 m_{e} c \alpha^2} \, ,
\end{equation}
is the Lyman-$\alpha$ rest wavelength.

The two-photon process $2s \rightarrow 1s$ proceeds through virtual atomic
states at a rate $\Lambda_{2s \rightarrow 1s} = 8.22458 \ {\rm s}^{-1}$
\cite{Goldman89}, and scales as \cite{Breit40Shapiro59}
\begin{equation}
\Lambda_{2s \rightarrow 1s} \propto \alpha^8 \, .
\label{eqn:twophoton}
\end{equation}

\begin{figure}
\centerline{\psfig{file=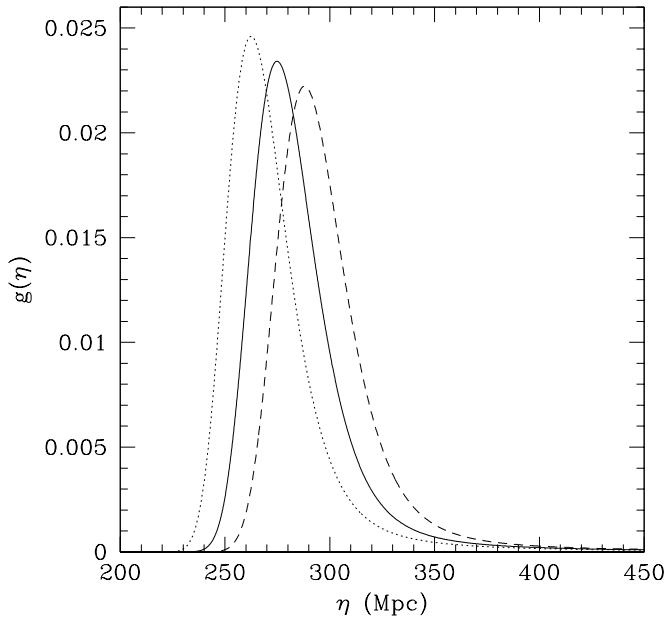,width=3.67in,angle=0}}
\caption{The probability distribution for the last scattering of a photon, the
visibility function, as a function of conformal time $\eta$ in the $\Lambda$CDM
model for $\varphi=(\alpha-\alpha_0)/\alpha_0=0.03$ (dotted), $\varphi=0$
(solid), and $\varphi=-0.03$ (dashed). }
\label{fig:vis}
\end{figure}

Eqs.~(\ref{eqn:recomb})--(\ref{eqn:twophoton}) account for
the $\alpha$ dependence of $x_e(t)$.  Along with the $\alpha$ dependence of
$\sigma_{T}$, this completely determines how $g(t)$ varies with
$\alpha$ in a given cosmology. While we expect a more complete calculation of
recombination, such as that in Ref.~{\cite{Seager00}}, may yield further
refinements to the $\alpha$ dependence, modifying
hydrogen recombination as described above is adequate for our purposes because
we are primarily concerned with the $\alpha$ dependence of the visibility
function at very high redshift where the simple calculation accurately captures
the physics.  Also, it was determined in Ref.~{\cite{Kap99}} that other
effects, such as modifications to the details of helium recombination, or the
cooling of baryons, are small compared to the effect of
variations of $\alpha$ on hydrogen recombination.

In this paper we work within the flat geometry $\Lambda$CDM cosmology with
baryon and matter densities $\Omega_{b}=0.05$ and $\Omega_{m}=0.30$, Hubble
parameter $h=0.72$, and spectral index $n=1$. Fig.~{\ref{fig:vis}} shows the
visibility function $g(\eta)={\rm exp}(-\tau)d\tau/d\eta$ plotted versus the
conformal time $\eta=\int dt/a$ for three different values of
$\varphi=({\alpha-\alpha_{0}})/{\alpha_{0}}$ where $\alpha_{0}=0.00729735
\simeq {1}/{137}$ is the value of the electromagnetic fine structure parameter
\cite{Mohr99}.  For positive values of $\varphi$  the visibility function is
narrower and peaks earlier, while for negative values of $\varphi$ the
visibility
function is broader and peaks later.  These effects impact directly the CMB
angular power spectrum because the peak of the visibility function
determines the physical distance to the last scattering surface, while the
width of the visibility function determines the thickness of the last
scattering surface.

\begin{figure}
\centerline{\psfig{file=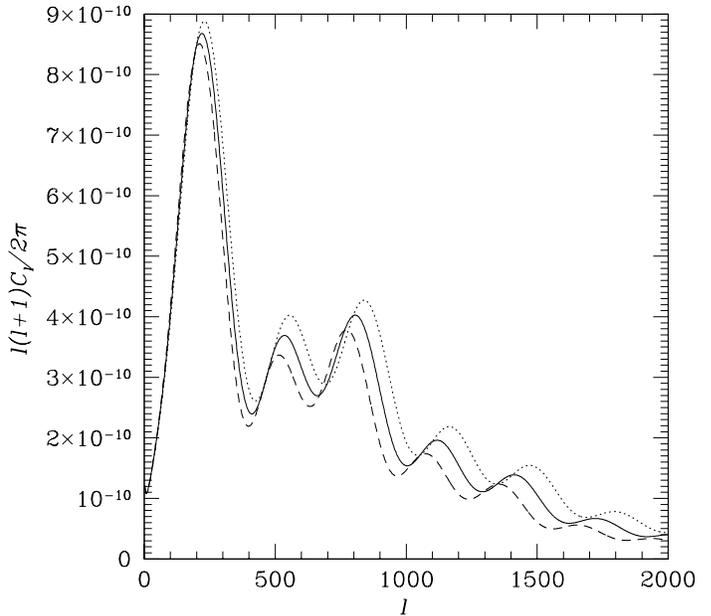,width=3.67in,angle=0}}
\caption{The angular power spectrum of CMB anisotropies for a spatially uniform
$\varphi=0.03$ (dotted), $\varphi=0$ (solid), and $\varphi=-0.03$ (dashed). }
\label{fig:cmb}
\end{figure}

Fig.~{\ref{fig:cmb}} shows the angular power spectrum of CMB anisotropies
calculated assuming a spatially homogeneous value of $\varphi$.  The dependence
of the
spectrum on $\varphi$ is easy to understand qualitatively.  For
positive $\varphi$ the
angular-diameter distance is larger, so the features are scaled
systematically to higher values of $l$.  The last-scattering surface is also
narrower so that small-scale (high-$l$) features are less damped due to photon
diffusion.   For negative $\varphi$ the opposite holds, the smaller
angular-diameter distance scales features to lower values of $l$ while the
broader last-scattering surface leads to more damping of power on small scales.
In the next two Sections we derive the changes in the angular power
spectrum due to spatial fluctuations in $\varphi$ between
causally disconnected regions of the Universe.

\section{Power Spectra}
\label{sec:PowSpect}
\subsection{CMB Power Spectra Fundamentals}

The CMB radiation is observed to be a nearly isotropic background of blackbody
radiation at a temperature of $T_{\rm CMB}=2.728 \pm 0.004 \, {\rm K}$
\cite{Fixsen96}. Anisotropies in the temperature are observed with a fractional
amplitude of $\sim10^{-5}$ \cite{Smoot92}, and in the polarization
with a fractional amplitude of $\sim10^{-6}$ \cite{Kovac02Kogut03}.
For a review of the physics of CMB anisotropies see, e.g., Ref.~{\cite{Hu96}.

The fundamental CMB anisotropy observables
are the Stokes parameters $(\Theta,Q,U,V)$, which
can be expressed in the Pauli
basis as (see, for example, Ref.~\cite{Bond87Kosowsky96})
\begin{equation}
{\bf P}(\vhat{n})=\Theta(\vhat{n}){\bf 1}+Q(\vhat{n}){\bf
\sigma_{3}}+U(\vhat{n}){\bf \sigma_{1}}+V(\vhat{n}){\bf \sigma_{2}} \, .
\end{equation}
In this expression,
\begin{equation}
\Theta(\vhat{n})=\frac{T(\vhat{n})-T_{\rm CMB}}{T_{\rm CMB}} \, ,
\end{equation}
denotes the fractional temperature anisotropy in a direction $\vhat{n}$, and
the remaining Stokes parameters are normalized to this quantity.  Here $Q$ and
$U$ describe independent linear polarization states, while $V$ describes
circular polarization.  Because circular polarization cannot be generated via
Thomson scattering, $V=0$ for the CMB.

It is convenient to introduce the quantities,
\begin{equation}
{_{\pm}A(\vhat{n})}=Q(\vhat{n}) \pm i U(\vhat{n}) \, ,
\end{equation}
which have the spin-2 transformation properties,
\begin{equation}
{_{\pm}A(\vhat{n})} \rightarrow {e}^{\mp 2 i \phi} {_{\pm}A(\vhat{n})} \, ,
\end{equation}
under a counterclockwise rotation of the coordinate axis by an angle $\phi$.

We can expand $\Theta(\vhat{n})$ and ${_{\pm}A(\vhat{n})}$
in normal modes as \cite{Zaldar97Kamion97Hu97},
\begin{equation}
\Theta(\vhat{n})=  \sum_{l=1}^{\infty} \sum_{m=-l}^{m=l} (-i)^{l}
\sqrt{\frac{4\pi}{2l+1}}\Theta_{lm} Y_{l}^{m}(\vhat{n}) \, ,
\end{equation}
and
\begin{equation}
{_{\pm}A(\vhat{n})}= \sum_{l=1}^{\infty} \sum_{m=-l}^{m=l} (-i)^{l}
\sqrt{\frac{4\pi}{2l+1}} \left( {_{\pm}A}_{lm}
\right)\left({_{\pm2}Y}_{l}^{m}(\vhat{n})\right) \, ,
\end{equation}
where
\begin{equation}
\Theta_{lm}=\int \frac{d^3\vect{k}}{(2\pi)^3} \Theta_{l}^{(m)}(k)
e^{i\vect{k}\cdot\vect{x}} \, ,
\end{equation}
and
\begin{equation}
{_{\pm}A}_{lm}=\int \frac{d^3\vect{k}}{(2\pi)^3} \left( {_{\pm}A}_{l}^{(m)}(k)
\right) e^{i\vect{k}\cdot\vect{x}} \, .
\end{equation}
Here ${_{s}Y}_{l}^{m}$ are the spin-$s$ weighted spherical harmonics
\cite{Newman66Goldberg66Thorne80}, with $Y_{l}^{m}={_{0}Y}_{l}^{m}$, and
$X_{l}^{(m)}(k)$ is the contribution to the angular
mode $X_{lm}$ from wave vectors of the primordial density field of magnitude
$k$.

It is conventional to write the polarization in terms of the moments of the
curl-free (scalar) configurations $E_{lm}$, and the moments of the
divergence-free (pseudo-scalar) configurations $B_{lm}$, where
\begin{equation}
{_{\pm}A}_{lm} = E_{lm} \pm i B_{lm} \, .
\end{equation}
We can then provide a complete description of an arbitrary CMB anisotropy field
using the moments $\Theta_{lm}$, $E_{lm}$, and $B_{lm}$.

The basic observables of the random fields $X(\vhat{n})$ are the
power spectra $C_{l}^{X\tilde{X}}$, defined by,
\begin{equation}
\<{X^{*}_{lm}\widetilde{X}_{l^{\prime}m^{\prime}}} =
\delta_{ll^{\prime}}\delta_{mm^{\prime}}C_{l}^{X\tilde{X}} \, ,
\end{equation}
where $X,\widetilde{X} \in \{\Theta,E,B\}$ and the angle brackets denote an
average over all realizations. Here,
\begin{equation}
C_{l}^{{X\tilde{X}}}=\frac{2}{\pi(2l+1)^2} \int \frac{dk}{k} \sum_{m=-2}^{2}
k^3 X_{l}^{(m)*}(k)\widetilde{X}_{l}^{(m)}(k) \, .
\label{eqn:allsky}
\end{equation}

A set of Gaussian random fields---and we expect $\{\Theta, E, B
\}$ to be Gaussian---are completely characterized by
their power spectra and cross-power spectra.
Because the pseudo-scalar $B$ has opposite parity to the scalars $\Theta$ and
$E$ the only non-vanishing
power spectra are $C_{l}^{\Theta\Theta}$, $C_{l}^{\Theta E}$, $C_{l}^{EE}$, and
$C_{l}^{BB}$.

For small patches of sky it is an excellent approximation to treat the sky as
flat and expand the field $X(\vhat{n})$ in spin-weighted Fourier modes rather
than spin-weighted spherical harmonics.  Thus we have (for example
Ref.~\cite{Hu00})
\begin{equation}
\Theta(\vhat{n})= \int \frac{d^2\vect{l}}{(2\pi)^2} \Theta(\vect{l})
e^{i\vect{l}\cdot\vhat{n}} \, ,
\label{eqn:thetafourier}
\end{equation}
\begin{align}
{_{\pm}A(\vhat{n})}= -\int \frac{d^2\vect{l}}{(2\pi)^2} {_{\pm}A(\vect{l})}
e^{\pm2i(\phi_{l}-\phi)} e^{i\vect{l}\cdot\vhat{n}} \, ,
\label{eqn:afourier}
\end{align}
and we again define $E$ and $B$ through
\begin{equation}
{_{\pm}A}(\vect{l}) = E(\vect{l}) \pm i B(\vect{l}) \, .
\end{equation}
In this notation the power spectra are defined by
\begin{equation}
\<{X(\vect{l})\widetilde{X}(\vect{l}^{\prime})}=(2\pi)^2
\delta^2(\vect{l}+\vect{l}^{\prime})C_{l}^{X\tilde{X}} \, .
\end{equation}
Unless otherwise noted, we work within this flat-sky approximation
for the remainder of the paper.

\subsection{Derivative Power Spectra}
\label{sec:derivpower}

How do the expressions for the power spectra change if we allow for spatial
fluctuations of $\alpha$? As a warmup we first consider a spatially uniform
variation, $\alpha=\alpha_0(1+\varphi)$, where $\varphi \ll 1$.

For a given primordial density field $\delta(\vect{x})$, the temperature and
polarization patterns, $\Theta(\vhat{n})$ and ${_{\pm}A}(\vhat{n})$, can be
calculated by solving the combined Einstein equations and radiative-transfer
equations, as well as the equations for the recombination history.  As
discussed above, this recombination history depends on $\alpha$.  Thus, the
temperature and polarization fields are implicitly functions of
$\varphi=(\alpha-\alpha_{0})/\alpha_{0}$. We can expand
$\Theta(\vhat{n})=\Theta(\vhat{n}; \varphi)$ and
${_{\pm}A(\vhat{n})}={_{\pm}A(\vhat{n}; \varphi)}$ in Taylor series about
$\varphi=0$,
\begin{eqnarray}
\Theta(\vhat{n})=
\Theta_{0}(\vhat{n}) + \pphi\Theta_{0}(\vhat{n})\varphi
+ \frac{1}{2}\pphi^2\Theta_{0}(\vhat{n})\varphi^2 + \cdots \, ,
\end{eqnarray}
\begin{align}
{_{\pm}A}(\vhat{n})=
{_{\pm}A}_{0}(\vhat{n}) + \pphi \left({_{\pm}A}_{0}(\vhat{n})\right)\varphi
&+ \frac{1}{2}\pphi^2 \left({_{\pm}A}_{0}(\vhat{n})\right)\varphi^2 \nonumber
\\ &+ \cdots \, .
\end{align}
Inverting Eqs.~({\ref{eqn:thetafourier}}) and ({\ref{eqn:afourier}}), we find
that
\begin{equation}
\Theta(\vect{l})=\int d^2\vhat{n} \,\Theta(\vhat{n})
e^{-i\vect{l}\cdot\vhat{n}} \, ,
\end{equation}
and
\begin{align}
{_{\pm}A(\vect{l})}= -\int {d^2\vhat{n}} \left({_{\pm}A(\vhat{n})}\right)
e^{\pm2i(\phi-\phi_{l})} e^{-i\vect{l}\cdot\vhat{n}} \, .
\end{align}
Thus, the $\varphi$ expansions can be written in $\vect{l}$ space as
\begin{equation}
\Theta(\vect{l}) = \Theta_{0}(\vect{l}) +
\pphi\Theta_{0}(\vect{l})\varphi +
\frac{1}{2}\pphi^2\Theta_{0}(\vect{l})\varphi^2 + \cdots \, ,
\end{equation}
\begin{equation}
{_{\pm}A}(\vect{l})={_{\pm}A}_{0}(\vect{l})+\pphi
\left({_{\pm}A}_{0}(\vect{l})\right)\varphi +
\frac{1}{2}\pphi^2\left({_{\pm}A}_{0}(\vect{l})\right)\varphi^2
+ \cdots \, ,
\end{equation}
and in fact for any field $X \in \{ \Theta, E, B \}$ we may write
\begin{equation}
X(\vect{l})=X_{0}(\vect{l})+\pphi
X_{0}(\vect{l})\varphi+\frac{1}{2}\pphi^2X_{0}(\vect{l})\varphi^2 + \cdots \, .
\end{equation}

To $O(\varphi^2)$ we then have,
\begin{align}
\langle  X(\vect{l})&\widetilde{X}(\vect{l'})  \rangle =
\<{ X_{0}(\vect{l})\widetilde{X}_{0}(\vect{l'}) } \nonumber
\\
&+\<{ X_{0}(\vect{l})\pphi \widetilde{X}_{0}(\vect{l'}) }\varphi
+ \<{ \pphi X_{0}(\vect{l})\widetilde{X}_{0}(\vect{l'}) }\varphi \nonumber
\\
&+ \frac{1}{2}\<{ X_{0}(\vect{l})\pphi ^2\widetilde{X}_{0}(\vect{l'})
}\varphi^2+ \frac{1}{2}\<{ \pphi ^2X_{0}(\vect{l})\widetilde{X}_{0}(\vect{l'})
}\varphi^2 \nonumber
\\
&+ \<{ \pphi X_{0}(\vect{l})\pphi \widetilde{X}_{0}(\vect{l'}) }\varphi^2,
\end{align}
and in terms of power spectra this becomes
\begin{align}
C_{l}^{{X\tilde{X}}}&=C_{l}^{{X_{0}\tilde{X}_{0}}} +
\left(C_{l}^{{X_{0}\partial \tilde{X}_{0}}}+C_{l}^{{\partial X_{0}
\tilde{X}_{0}}}\right)\varphi \nonumber
\\ &+
\left(\frac{1}{2}C_{l}^{X_{0} \partial^2\tilde{X}_{0}} +
\frac{1}{2}C_{l}^{\partial^2X_{0} \tilde{X}_{0}}+C_{l}^{\partial
X_{0}\partial \tilde{X}_{0}}\right)\varphi^2.
\end{align}

Since we are for the time being assuming a spatially uniform value of $\varphi$
we may also write
\begin{align}
C_{l}^{{X\tilde{X}}}=\left. C_{l}^{{X\tilde{X}}}\right|_{0} + \left. \pphi
C_{l}^{{X\tilde{X}}}\right|_{0}\varphi +
\frac{1}{2}\left.\pphi^2C_{l}^{X\tilde{X}} \right|_{0}\varphi^2.
\end{align}
This allows us to make the identifications
\begin{align}
\left. C_{l}^{{X\tilde{X}}}\right|_{0} &=  C_{l}^{{X_{0}\tilde{X}_{0}}} \, ,\nonumber \\
\left. \pphi C_{l}^{{X\tilde{X}}}\right|_{0} &=  C_{l}^{{X_{0}\partial
\tilde{X}_{0}}}+C_{l}^{{\partial X_{0} \tilde{X}_{0}}} \, ,\nonumber \\
\left. \pphi^2C_{l}^{X\tilde{X}} \right|_{0} &=
C_{l}^{X_{0}\partial^2\tilde{X}_{0}}+C_{l}^{\partial^2X_{0}
\tilde{X}_{0}}+2C_{l}^{\partial X_{0}\partial \tilde{X}_{0}} \, .
\end{align}

\begin{figure}
\centerline{\psfig{file=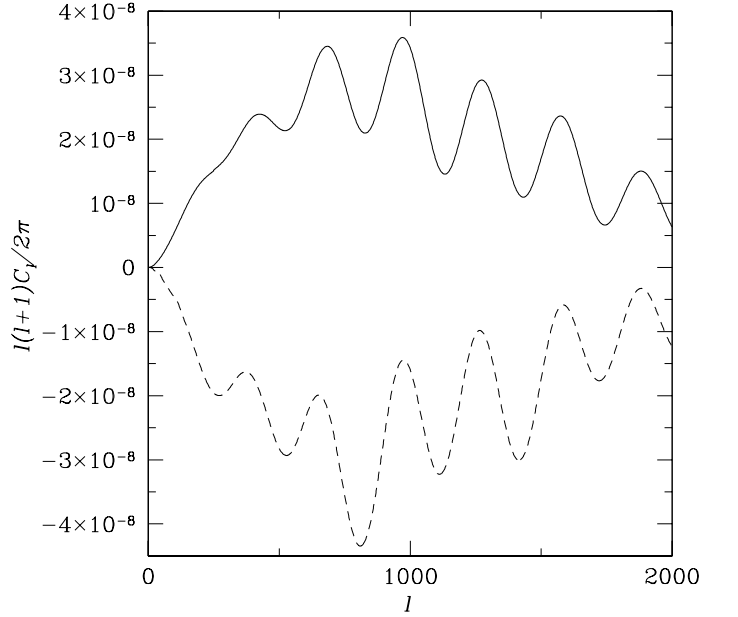,width=3.67in,angle=0}}
\caption{The $C_{l}^{\partial \Theta_{0} \partial \Theta_{0}}$ (solid) and
$C_{l}^{\Theta_{0} \partial^2 \Theta_{0}}$ (dashed) derivative power spectra.}
\label{fig:TT_deriv}
\end{figure}

\begin{figure}
\centerline{\psfig{file=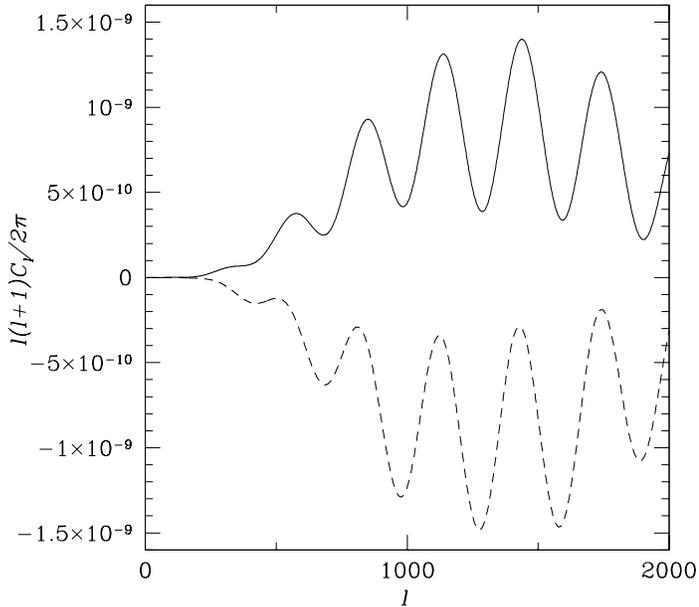,width=3.67in,angle=0}}
\caption{The $C_{l}^{\partial E_{0} \partial E_{0}}$ (solid) and $C_{l}^{E_{0}
\partial^2 E_{0}}$ (dashed) derivative power spectra.}
\label{fig:EE_deriv}
\end{figure}

\begin{figure}
\centerline{\psfig{file=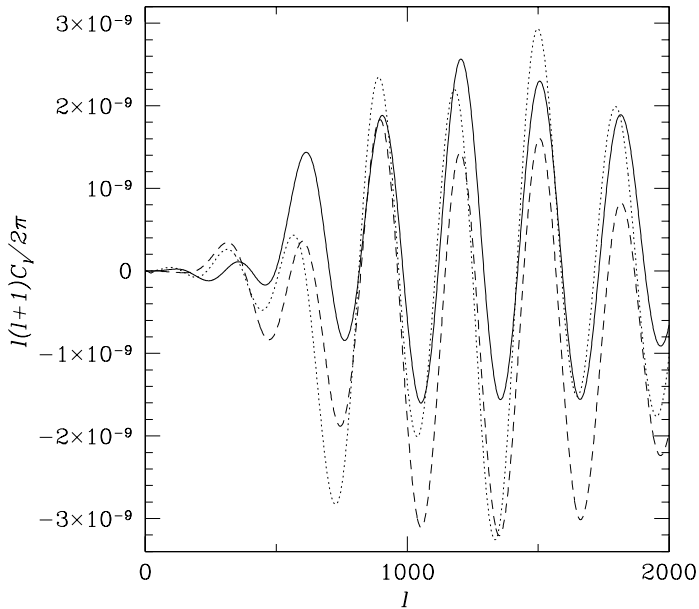,width=3.67in,angle=0}}
\caption{The $C_{l}^{\partial \Theta_{0} \partial E_{0}}$ (solid),
$C_{l}^{\Theta_{0} \partial^2 E_{0}}$ (dashed) and $C_{l}^{E_{0} \partial^2
\Theta_{0}}$ (dotted) derivative power spectra.}
\label{fig:TE_deriv}
\end{figure}
\noindent These identifications make it clear how to calculate the individual
`derivative' power spectra.  We just differentiate Eq.~(\ref{eqn:allsky}) with
respect to $\varphi$, evaluate the expression at $\varphi=0$, and pick off the
terms with the requisite structure.

For example,
\begin{align}
C_{l}^{{\partial X_{0}\partial \tilde{X}_{0}}}&= \frac{2}{\pi(2l+1)^2} \int
\frac{dk}{k} \nonumber \\
&\times \left. \sum_{m=-2}^{2} k^3 \pphi X_{l}^{(m)*}(k)\pphi
\widetilde{X}_{l}^{(m)}(k) \right|_{\varphi=0} \, ,
\end{align}
where factors like $\pphi \widetilde{X}_{l}^{(m)}(k)$ can be
calculated numerically or directly from first principles using the
expressions for the $\widetilde{X}_{l}^{(m)}(k)$ derived in, for example,
Ref.~{\cite{Hu98}}.  Because they are used in subsequent calculations we have
created a modified version of the code {\tt CMBFAST} \cite{Seljak96} that can
compute these derivative power spectra. (See Figs.~3--5.)

\section{Spatial Variations of {\large $\alpha$}}

We now consider the effects of spatial variations of $\alpha$ (parametrized by
$\varphi$) between different causally
disconnected regions of the Universe.

First suppose that there were \emph{no} density fluctuations, but spatial
variations of $\alpha$.  In that case, photons from different points on the sky
would be last scattered at different cosmological times, but they would all
still have the same frequency when observed by us.  However, if there are
density fluctuations, the manner in which they are imprinted on the CMB depends
on the value of $\alpha$, as discussed above.  Thus, if there are spatial
variations in $\alpha$, the power spectra (or two-point correlation functions)
will vary from one place on the sky to another.  This implies that the stochasticity of the spatial
variations in $\alpha$ induce non-Gaussianity in the CMB quantified by non-zero
(connected) higher order correlation functions (trispectra and perhaps
bispectra).  It also implies a correction to the mean power spectrum (i.e.,
that measured by mapping regions of the sky that contain many coherence regions
of $\alpha$), as well as  the introduction of a non-zero curl in the
polarization.  All of these effects are analogous to similar effects induced by
weak lensing of the CMB.  The only difference is that in our case, the
temperature and polarization patterns are modulated by a variable $\alpha$,
rather than lensing by an intervening density field along the line of sight.

In this Section, we first calculate the modified power spectra
$C_{l}^{\Theta\Theta}$, $C_{l}^{\Theta E}$, $C_{l}^{EE}$, and $C_{l}^{BB}$.  We
then determine the form of the higher order correlations (bispectra and
trispectra) in the next two Sections.

\subsection{Observable modes in the presence of $\varphi$ fluctuations}

We assume that at a given position $\vhat{n}$ at the surface of last scatter,
the value of $\alpha$ is $\alpha(\vhat{n})=\alpha_0[1+\varphi(\vhat{n})]$.
Here we treat $\varphi(\vhat{n})$ as a random field with angular power spectrum
$\<{ \varphi(\vect{l})\varphi(\vect{l'}) } =
(2\pi)^2\delta^2(\vect{l}+\vect{l'})C_{l}^{{\varphi\varphi}}$ in the flat-sky
approximation.

We assume that the surface of last scatter is much thinner than the spatial
correlation length
of $\varphi$, and that in a given direction $\alpha$ is constant throughout
recombination.  We also assume that the dynamics responsible for the variations
of $\varphi$ have a negligible effect on the perturbation evolution so that
the sole effect of variations of $\varphi$ are a modification of the
microphysics.  We will discuss the validity of these assumptions later.

Again we expand our fields  $\Theta(\vhat{n})=\Theta(\vhat{n}; \varphi)$
and ${_{\pm}A}(\vhat{n})={_{\pm}A}(\vhat{n}; \varphi)$ in a Taylor series about
$\varphi=0$ as
\begin{align}
\Theta(\vhat{n})=
\Theta_{0}(\vhat{n}) &+ \pphi \Theta_{0}(\vhat{n})\varphi(\vhat{n}) \nonumber
\\
&+ \frac{1}{2}\pphi^2 \Theta_{0}(\vhat{n})\varphi^2(\vhat{n}),
\end{align}
\begin{align}
{_{\pm}A}(\vhat{n})=
{_{\pm}A}_{0}(\vhat{n}) &+ \pphi
\left({_{\pm}A}_{0}(\vhat{n})\right)\varphi(\vhat{n}) \nonumber
\\
&+ \frac{1}{2}\pphi^2 \left( {_{\pm}A}_{0}(\vhat{n})\right)\varphi^2(\vhat{n}),
\end{align}
where $\varphi(\vhat{n})$ is now a function of position.

By taking the Fourier transform of $\Theta(\vhat{n})$ we find
\begin{equation}
\Theta(\vect{l})=\Theta_{0}(\vect{l})+\left[(\pphi\Theta_{0}) \conv
\varphi\right](\vect{l})+\frac{1}{2}\left[(\pphi^2\Theta_{0}) \conv \varphi
\conv \varphi\right](\vect{l})\,
\label{eqn:theta}
\end{equation}
where
\begin{equation}
\left[X \conv \psi\right](\vect{l}) = \int \frac{d^2 \vect{l'}}{(2\pi)^2}
X(\vect{l'})\psi(\vect{l}-\vect{l'}) \,
\end{equation}
is the convolution of two fields $X$ and $\psi$,
and
\begin{equation}
\left[X \conv \psi \conv \lambda \right](\vect{l}) = \int \frac{d^2
\vect{l'}}{(2\pi)^2} \left[ X \conv \psi
\right](\vect{l'})\lambda(\vect{l}-\vect{l'}) \,
\end{equation}
is the double convolution of three fields $X$, $\psi$, and $\lambda$.

Similarly, taking linear combinations of the Fourier transforms of
${_{\pm}A}(\vhat{n})$ we find that
\begin{align}
E(\vect{l})=E_{0}(\vect{l})&+\left[(\pphi E_{0}) \convc
\varphi\right](\vect{l})-\left[(\pphi B_{0}) \convs \varphi\right](\vect{l})
\nonumber \\
&+\frac{1}{2}\left[(\pphi^2 E_{0}) \convc \varphi \conv
\varphi\right](\vect{l})\nonumber \\
&-\frac{1}{2}\left[(\pphi^2 B_{0}) \convs \varphi \conv
\varphi\right](\vect{l}),
\label{eqn:e}
\end{align}
and
\begin{align}
B(\vect{l})=B_{0}(\vect{l})&+\left[(\pphi B_{0}) \convc
\varphi\right](\vect{l})+\left[(\pphi E_{0}) \convs \varphi\right](\vect{l})
\nonumber \\
&+\frac{1}{2}\left[(\pphi^2 B_{0}) \convc \varphi \conv
\varphi\right](\vect{l})\nonumber \\
&+\frac{1}{2}\left[(\pphi^2 E_{0}) \convs \varphi \conv
\varphi\right](\vect{l}),
\label{eqn:b}
\end{align}
where
\begin{equation}
\left[X \convc \psi\right](\vect{l}) = \int \frac{d^2 \vect{l'}}{(2\pi)^2}
\cos{(2\phi_{l'}
)}X(\vect{l'})\psi(\vect{l}-\vect{l'}) \, ,
\end{equation}
and
\begin{equation}
\left[X \convs \psi\right](\vect{l}) = \int \frac{d^2 \vect{l'}}{(2\pi)^2}
\sin{(2\phi_{l'}
)}X(\vect{l'})\psi(\vect{l}-\vect{l'}) \, ,
\end{equation}
are the even- and odd-parity spin-2 weighted convolutions of $X$ and $\psi$
respectively.

Examining these expressions we find that a given mode $X(\vect{l})$ receives
corrections due to the combination of modes $\{ \pphi^{n}X_{0}(\vect{l}_0),
\varphi(\vect{l}_1), \varphi(\vect{l}_2), \dots , \varphi(\vect{l}_n) \}$ such
that $\sum_{i=0}^{n} \vect{l}_i=\vect{l}$.  Furthermore, the $E$ and $B$ modes
mix so that, for example, the mode $B(\vect{l})$ can be induced by the
combinations of modes $\{ \pphi^{n}E_{0}(\vect{l}_0), \varphi(\vect{l}_1),
\varphi(\vect{l}_2), \dots , \varphi(\vect{l}_n) \}$ such that $\sum_{i=0}^{n}
\vect{l}_i=\vect{l}$.  These effects modify the angular power spectra of CMB
anisotropies, and introduce higher-order connected (non-Gaussian) correlation
functions.

\subsection{The $\Theta\Theta$ Power Spectrum}

Using Eq.~(\ref{eqn:theta}) we find that the expansion for the two point
correlation function (in Fourier space) is
\begin{align}
\langle \Theta(&\vect{l})\Theta(\vect{l'}) \rangle =
\<{ \Theta_{0}(\vect{l})\Theta_{0}(\vect{l'}) } \nonumber
\\
&+ \<{ \Theta_{0}(\vect{l})\left[\pphi\Theta_{0}\conv\varphi\right] (\vect{l'})
}
+ \<{ \left[\pphi\Theta_{0}\conv\varphi\right](\vect{l})\Theta_{0}(\vect{l'}) }
\nonumber
\\
&+ \frac{1}{2}\<{
\Theta_{0}(\vect{l})\left[\pphi^2\Theta_{0} \conv
\varphi\conv\varphi\right](\vect{l'}) }
\nonumber \\
&+ \frac{1}{2}\<{
\left[\pphi^2\Theta_{0}\conv\varphi\conv\varphi\right]
(\vect{l})\Theta_{0}(\vect{l'}) } \nonumber
\\
&+ \<{
\left[\pphi\Theta_{0}\conv\varphi\right](\vect{l})
\left[\pphi\Theta_{0}\conv\varphi\right](\vect{l'}) }
\end{align}
to $O(\varphi^2)$.  In the above expression and what follows we adopt the
convention that the differential operators $\partial_{\varphi}$ act only the
the field immediately following them.

We assume that $\Theta_{0}$ and $\varphi$ are zero-mean Gaussian random fields
without higher-order connected correlators. By writing out the convolutions and
Wick expanding the correlators it is easy to verify that correlators involving
an odd number of fields vanish, and so there are no corrections to first
order in $\varphi$.  It is also straightforward to verify that
\begin{align}
\langle &\Theta_{0}(\vect{l})[\pphi^2 \Theta_{0}\conv\varphi\conv
\varphi](\vect{l'})  \rangle = \langle [ \pphi^2 \Theta_{0}\conv\varphi\conv
\varphi](\vect{l})\Theta_{0}(\vect{l'})  \rangle \nonumber \\
& =(2\pi)^2 \delta^2(\vect{l}+\vect{l'})
\left[\sigma^{(\varphi\varphi)}C_{l}^{\Theta_{0}\partial^2\Theta_{0}}
+ 2\sigma^{(\varphi\partial^2\Theta_{0})}C_{l}^{\Theta_{0}\varphi}\right] \, ,
\end{align}
where
\begin{align}
\sigma^{(\psi\lambda)}=\int \frac{d^2 \vect{l}}{(2\pi)^2} C_{l}^{\psi\lambda}
\end{align}
is the covariance between two fields $\psi$ and $\lambda$.  Similarly we can
show that
\begin{align}
\langle
\left[\pphi\Theta_{0}\conv\varphi\right](\vect{l})
\left[\pphi\Theta_{0}\conv\varphi\right](\vect{l'}) \rangle
\hspace{-0.5cm} & \nonumber \\
=(2\pi)^2 \delta^2(\vect{l}+\vect{l'})
\{ &\left[C^{\varphi\varphi}\conv
C^{\partial\Theta_{0}\partial\Theta_{0}}\right]_{l}
\nonumber \\
+ & \left[C^{\partial\Theta_{0}\varphi}\conv
C^{\partial\Theta_{0}\varphi}\right]_{l} \} \, ,
\end{align}
where we have dropped terms that contribute only when $\vect{l}=0$.

Collecting all terms we
find that to leading order the average power spectrum including
fluctuations in $\varphi$ is
\begin{align}
C_{l}^{{\Theta\Theta}} = C_{l}^{{\Theta_{0}\Theta_{0}}} &+
\sigma^{(\varphi\varphi)}C_{l}^{\Theta_{0}\partial^2\Theta_{0}}
+ 2\sigma^{(\varphi\partial^2\Theta_{0})}C_{l}^{\Theta_{0}\varphi} \nonumber \\
&+\left[C^{\varphi\varphi}\conv
C^{\partial\Theta_{0}\partial\Theta_{0}}\right]_{l}
+ \left[C^{\partial\Theta_{0}\varphi}\conv
C^{\partial\Theta_{0}\varphi}\right]_{l} .
\end{align}

\begin{figure}
\centerline{\psfig{file=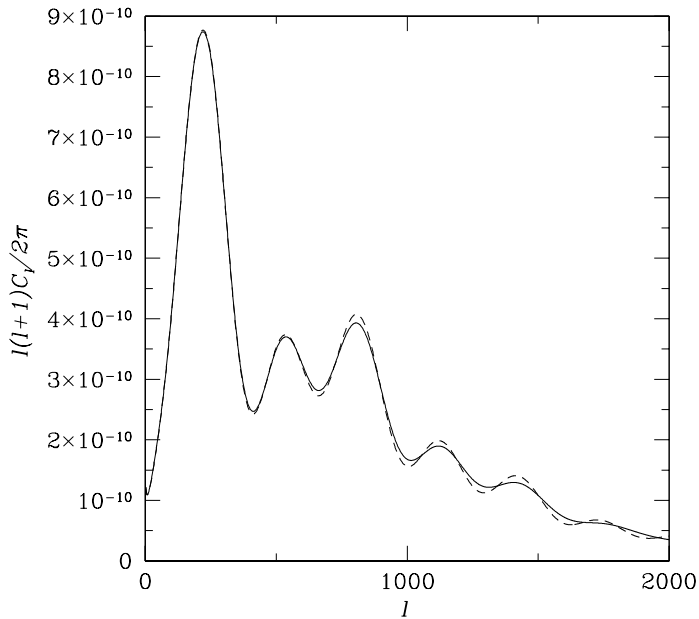,width=3.67in,angle=0}}
\caption{The solid curve shows the $\Theta \Theta$ power spectrum for the
Gaussian correlation function for $\varphi$ with angular correlation scale
$\theta_{c}=1^{\circ}$ and variance $\sigma^{(\varphi \varphi)}=9\times
10^{-4}$ (fluctuations in $\varphi$ at the 3\% level).  The dashed curve shows
the power spectrum without
fluctuations in $\varphi$.}
\label{fig:TT_powspect}
\end{figure}

\noindent Note that the corrections to
$C_{l}^{\Theta_{0}\Theta_{0}}$ involve couplings between the
derivative power spectra, which are calculated as described in
Section~{\ref{sec:derivpower}}, and $C_{l}^{\varphi\varphi}$ and
the various cross power spectra, which are specified by the
model that generates spatial variations in $\varphi$.

In general $\Theta_{0}$ and $\varphi$ may be correlated if, for
instance, they are generated by a common mechanism or if they
are strongly coupled through evolution equations.  We discuss in
Section~{\ref{sec:ParticleTheory} why we do not expect the
latter source of correlations to be important as long as the
energy in the $\varphi$ field and its fluctuations are small
compared to the dominant radiation and matter perturbations.  In
the case where $\Theta_{0}$ and $\varphi$ have no
cross-correlation, the
expression simplifies to
\begin{align}
C_{l}^{{\Theta\Theta}} = C_{l}^{{\Theta_{0}\Theta_{0}}} &+
\sigma^{(\varphi\varphi)}C_{l}^{\Theta_{0}\partial^2\Theta_{0}}
+\left[C^{\varphi\varphi}\conv
C^{\partial\Theta_{0}\partial\Theta_{0}}\right]_{l} .
\end{align}

The result for $C_{l}^{\Theta\Theta}$ will of course depend on
$C_{l}^{\varphi\varphi}$.  To illustrate we consider a simple model in which
$\varphi$ is highly correlated on angular scales smaller than the correlation
angle $\theta_{c}$, and uncorrelated on larger scales. (Below we discuss a
physical model that may produce such a correlation function.) We thus have,
\begin{align}
\<{ \varphi(0)\varphi(\theta)} =
&\sigma^{(\varphi\varphi)}{e}^{-({\theta}/{\theta_c})^2} \, .
\end{align}

In the flat-sky approximation we have
\begin{equation}
C_{l}^{{\varphi\varphi}} = \int d^2\theta \, \<{ \varphi(0)\varphi(\theta) }
e^{-i\vect{l}\cdot\mbox{\boldmath \scriptsize $\theta$}} \, ,
\end{equation}
which implies that
\begin{equation}
C_{l}^{\varphi\varphi} = \pi \theta_{c}^2 \sigma^{(\varphi\varphi)}
e^{-\frac{1}{4}l^2 \theta_c^2} \, .
\label{eqn:phipower}
\end{equation}

In this case the average $\Theta \Theta$ power spectrum is
\begin{align}
C_{l}^{{\Theta\Theta}} = &C_{l}^{{\Theta_{0}\Theta_{0}}}
+ \sigma^{(\varphi\varphi)}\left[ C_{l}^{\Theta_{0}\partial^2\Theta_{0}}
 \right. \nonumber \\
&+ \frac{\theta_c^2}{2}\int dl' l' e^{-\frac{1}{4}(l^2+{l'}^2)\theta_c^2} I_{0}
\left(\frac{\theta_{c}^2}{2}l l' \right) \left.
C_{l'}^{\partial\Theta_{0}\partial\Theta_{0}} \right] ,
\end{align}
where $I_{n}$ is the $n^{\rm th}$-order modified Bessel function of the first
kind.  We show this average power spectrum in Fig.~{\ref{fig:TT_powspect}.
The main effect of $\varphi$ fluctuations is to reduce the amplitude of
oscillatory features in the damping tail.  This effect can be understood by
noting that patches of the sky with different values of $\varphi$ will have
different power spectra, and that, as can be seen in Fig.~{\ref{fig:cmb}, the
location of of the
peaks in the damping tail of these power spectra shift as $\varphi$ changes.
These patch power spectra add incoherently, so that the amplitude of the
oscillatory component of the average power spectrum is reduced.  This is the
same type of effect as in weak gravitational lensing \cite{ZarSelLens}. As we
will see, this is the predominant effect in the other power spectra as well.

\subsection{The $\Theta E$ Power Spectrum}

Using Eqs.~(\ref{eqn:theta}) and (\ref{eqn:e}), and immediately dropping the
vanishing correlators involving an odd number of fields, we find that the
expansion for the power spectrum is
\begin{align}
\langle \Theta(\vect{l})E(\vect{l'}) \rangle &=
\<{ \Theta_{0}(\vect{l})E_{0}(\vect{l'}) } \nonumber
\\
&+ \frac{1}{2}\<{ \Theta_{0}(\vect{l})\left[\pphi^2
E_{0}\convc\varphi\conv\varphi \right](\vect{l'}) }  \nonumber
\\
&- \frac{1}{2}\<{ \Theta_{0}(\vect{l})\left[\pphi^2
B_{0}\convs\varphi\conv\varphi\right](\vect{l'}) } \nonumber
\\
&+ \<{ \left[\pphi\Theta_{0}\conv\varphi\right](\vect{l})\left[\pphi
E_{0}\convc\varphi\right](\vect{l'}) } \nonumber
\\
&- \<{ \left[\pphi\Theta_{0}\conv\varphi\right](\vect{l})\left[\pphi
B_{0}\convs\varphi\right](\vect{l'}) } \nonumber
\\
&+ \frac{1}{2}\<{
\left[\pphi^2\Theta_{0}\conv\varphi\conv\varphi\right](\vect{l})
E_{0}(\vect{l'})}
\end{align}
to $O(\varphi^2)$.

\begin{figure}
\centerline{\psfig{file=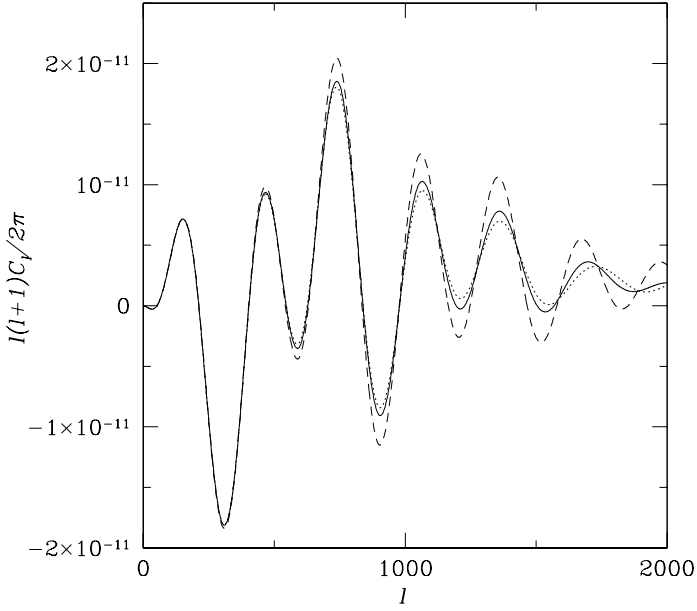,width=3.67in,angle=0}}
\caption{The $\Theta E$ power spectra for the Gaussian correlation function for
$\varphi$ with $\sigma^{(\varphi \varphi)}=9\times 10^{-4}$ and
$\theta_{c}=1^{\circ}$ (solid) and $\theta_{c}=2^{\circ}$ (dotted). The dashed
line shows the power spectrum without fluctuations in $\varphi$.}
\label{fig:TE_powspect}
\end{figure}

By writing out the convolutions, Wick expanding the correlators, and noting
that terms involving $B_{0}$ vanish due to parity, we find that the
non-vanishing
terms are
\begin{align}
\langle &\Theta_{0}(\vect{l})\left[\pphi^2 E_{0}\conv\varphi\conv
\varphi\right](\vect{l'})  \rangle \nonumber \\
& =(2\pi)^2 \delta^2(\vect{l}+\vect{l'})
\left[\sigma^{(\varphi\varphi)}C_{l}^{\Theta_{0}\partial^2 E_{0}}
+ 2\sigma^{(\varphi\partial^2 E_{0})}C_{l}^{\Theta_{0}\varphi}\right] \, ,
\end{align}
\begin{align}
\langle & \left[\pphi^2 \Theta_{0}\conv\varphi\conv \varphi\right](\vect{l})
E_{0}(\vect{l'})  \rangle \nonumber \\
& =(2\pi)^2 \delta^2(\vect{l}+\vect{l'})
\left[\sigma^{(\varphi\varphi)}C_{l}^{E_{0}\partial^2 \Theta_{0}}
+ 2\sigma^{(\varphi\partial^2 \Theta_{0})}C_{l}^{E_{0}\varphi}\right] \, ,
\end{align}
\begin{align}
\langle \left[\pphi\Theta_{0}\conv\varphi\right](\vect{l})\left[\pphi
E_{0}\convc\varphi\right](\vect{l'}) \rangle \hspace{-0.5cm} &\nonumber \\
 =(2\pi)^2 \delta^2(\vect{l}+\vect{l'}) \{ &
\left[C^{\partial\Theta_{0}\partial E_{0}} \convc C^{\varphi\varphi}\right]_{l}
\nonumber \\
+ & \left[C^{\partial E_{0}\varphi}\convc
C^{\partial\Theta_{0}\varphi}\right]_{l} \} \, .\end{align}

Collecting these terms we
find that, to leading order, the average cross power-spectrum including
fluctuations in $\varphi$ is
\begin{align}
C_{l}^{{\Theta E}} = C_{l}^{{\Theta_{0} E_{0}}} &+ \sigma^{(\varphi\varphi)}
\left[\frac{1}{2}C_{l}^{\Theta_{0}\partial^2 E_{0}} +
\frac{1}{2}C_{l}^{E_{0}\partial^2 \Theta_{0}} \right] \nonumber \\
&+ \sigma^{(\varphi\partial^2 E_{0})}C_{l}^{\Theta_{0}\varphi} +
\sigma^{(\varphi\partial^2 \Theta_{0})}C_{l}^{E_{0}\varphi} \nonumber \\
&+\left[C^{\partial\Theta_{0}\partial E_{0}}\convc C^{\varphi
\varphi}\right]_{l}
+ \left[C^{\partial E_{0}\varphi}\convc
C^{\partial\Theta_{0}\varphi}\right]_{l} \, .
\end{align}

If $\varphi$ has no correlation with the density field, the expression
simplifies to
\begin{align}
C_{l}^{{\Theta E}} = C_{l}^{{\Theta_{0} E_{0}}} &+ \sigma^{(\varphi\varphi)}
\left[\frac{1}{2}C_{l}^{\Theta_{0}\partial^2 E_{0}} +
\frac{1}{2}C_{l}^{E_{0}\partial^2 \Theta_{0}} \right] \nonumber \\
&+\left[C^{\partial\Theta_{0}\partial E_{0}}\convc C^{\varphi
\varphi}\right]_{l} \, .
\end{align}

Inserting the power spectrum from Eq.~(\ref{eqn:phipower}), we obtain the
expression
\begin{align}
C_{l}^{{\Theta E}} = &C_{l}^{{\Theta_{0} E_{0}}}
+ \sigma^{(\varphi\varphi)}\left[ \frac{1}{2}C_{l}^{\Theta_{0}\partial^2 E_{0}}
+ \frac{1}{2}C_{l}^{E_{0}\partial^2\Theta_{0}}
 \right. \nonumber \\
&+ \frac{\theta_c^2}{2}\int dl' l' e^{-\frac{1}{4}(l^2+{l'}^2)\theta_c^2} I_{2}
\left(\frac{\theta_{c}^2}{2}l l' \right) \left.
C_{l'}^{\partial\Theta_{0}\partial\Theta_{0}} \right] \, .
\end{align}
We show this average power spectrum in Fig~\ref{fig:TE_powspect}.  The major
effect of $\varphi$ fluctuations on the $\Theta E$ power spectrum is to
reduce the peak amplitudes on small scales.

\subsection{The $EE$ Power Spectrum}

Using Eq.~(\ref{eqn:e}), and dropping the correlators involving an odd number
of fields and those that vanish due to parity, we find that the
expansion for the power spectrum is
\begin{align}
\langle E(&\vect{l})E(\vect{l'}) \rangle =
\<{ E_{0}(\vect{l})E_{0}(\vect{l'}) } \nonumber
\\
&+ \frac{1}{2}\<{
E_{0}(\vect{l})\left[\pphi^2E_{0}\convc\varphi\conv\varphi\right](\vect{l'}) }
\nonumber
\\
&+ \frac{1}{2}\<{
\left[\pphi^2E_{0}\convc\varphi\conv\varphi\right](\vect{l})E_{0}(\vect{l'}) }
\nonumber
\\
&+ \<{ \left[\pphi E_{0}\convc\varphi\right](\vect{l})\left[\pphi
E_{0}\convc\varphi\right](\vect{l'}) }
\nonumber \\
&+ \<{ \left[\pphi B_{0}\convs\varphi\right](\vect{l})\left[\pphi
B_{0}\convs\varphi\right](\vect{l'}) }
\end{align}
to $O(\varphi^2)$.

\begin{figure}
\centerline{\psfig{file=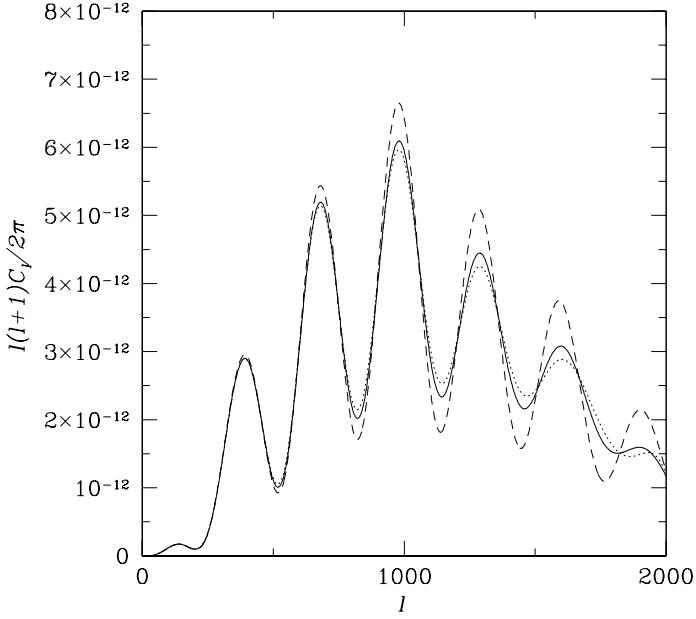,width=3.67in,angle=0}}
\caption{The $E E$ power spectra for the Gaussian correlation function for
$\varphi$ with $\sigma^{(\varphi \varphi)}=9\times 10^{-4}$ and
$\theta_{c}=1^{\circ}$ (solid) and $\theta_{c}=2^{\circ}$ (dotted). The dashed
line shows the power spectrum without fluctuations in $\varphi$.}
\label{fig:EE_powspect}
\end{figure}

After evaluating the convolutions and Wick expanding these correlators we find
\begin{align}
\langle & E_{0}(\vect{l})\left[\pphi^2  E_{0}\convc\varphi\conv
\varphi\right](\vect{l'})  \rangle = \langle \left[\pphi^2
E_{0}\convc\varphi\conv \varphi\right](\vect{l'}) E_{0}(\vect{l})  \rangle
\nonumber \\
& =(2\pi)^2 \delta^2(\vect{l}+\vect{l'}) \left[\sigma^{(\varphi\varphi)}C_{l}^{
E_{0}\partial^2 E_{0}}
+ 2\sigma^{(\varphi\partial^2 E_{0})}C_{l}^{ E_{0}\varphi}\right] \, ,
\end{align}
\begin{align}
\langle &\left[\pphi E_{0}\convc\varphi\right](\vect{l})\left[\pphi
E_{0}\convc\varphi\right](\vect{l'}) \rangle \nonumber \\
& =(2\pi)^2 \delta^2(\vect{l}+\vect{l'})\int \frac{d^2 \vect{l}_1}{(2\pi)^2}
\left\{ \cos^2{(2\phi_{l_1})}C_{l_1}^{\partial E_{0}\partial E_{0}}
C_{|\vect{l}-\vect{l}_1|}^{\varphi\varphi}  \right. \nonumber \\
& + \cos{(2\phi_{l_1})}\left[\frac{2 {l_1}^2 \sin^2{(\phi_{l_1})}
}{|\vect{l}-\vect{l}_1|^2} -1\right] \left. C_{l_1}^{\partial
E_{0}\varphi}C_{|\vect{l}-\vect{l}_1|}^{\partial E_{0}\varphi} \right\} \, ,
\end{align}
\begin{align}
\langle &\left[\pphi B_{0}\convs\varphi\right](\vect{l})\left[\pphi
B_{0}\convs\varphi\right](\vect{l'}) \rangle \nonumber \\
& =(2\pi)^2 \delta^2(\vect{l}+\vect{l'})\int \frac{d^2 \vect{l}_1}{(2\pi)^2}
 \sin^2{(2\phi_{l_1})}C_{l_1}^{\partial B_{0}\partial B_{0}}
C_{|\vect{l}-\vect{l}_1|}^{\varphi\varphi}  \, .
\end{align}
Collecting these terms we
find that
\begin{align}
&C_{l}^{EE} = C_{l}^{E_{0}E_{0}} + \sigma^{(\varphi\varphi)}
C_{l}^{E_{0}\partial^2 E_{0}}
+ 2\sigma^{(\varphi\partial^2 E_{0})}C_{l}^{ E_{0}\varphi} \nonumber \\
&+ \int \frac{d^2 \vect{l'}}{(2\pi)^2} \left\{
[ \cos^2{(2\phi_{l'})}C_{l'}^{\partial E_{0}\partial E_{0}} \right.
\nonumber \\
&+ \left.\sin^2{(2\phi_{l'})}C_{l'}^{\partial B_{0}\partial B_{0}} ]
C_{|\vect{l}-\vect{l'}|}^{\varphi\varphi} \right.
\nonumber \\
&+ \cos{(2\phi_{l'})}\left[\frac{2 {l'}^2 \sin^2{(\phi_{l'})}
}{|\vect{l}-\vect{l'}|^2} -1\right] \left. C_{l'}^{\partial
E_{0}\varphi}C_{|\vect{l}-\vect{l'}|}^{\partial E_{0}\varphi} \right\} \, ,
\end{align}
is the expression for the average $E$-mode power spectrum
including fluctuations in $\varphi$.

If $E_{0}$ and $\varphi$ are uncorrelated, the expression simplifies to
\begin{align}
C_{l}^{EE} = C_{l}^{E_{0}E_{0}} + \sigma^{(\varphi\varphi)}
C_{l}^{E_{0}\partial^2 E_{0}} \hspace{-0.72cm} &
\nonumber \\
+ \int \frac{d^2 \vect{l'}}{(2\pi)^2}
[ &\cos^2{(2\phi_{l'})}C_{l'}^{\partial E_{0}\partial E_{0}}
\nonumber \\
+ &\sin^2{(2\phi_{l'})}C_{l'}^{\partial B_{0}\partial B_{0}} ]
C_{|\vect{l}-\vect{l'}|}^{\varphi\varphi}
\, .
\end{align}

For the $\varphi$ power spectrum in Eq~(\ref{eqn:phipower}), dropping the
negligible primordial $B$-mode term, we obtain the expression
\begin{align}
C_{l}^{{E E}} = C_{l}^{{E_{0} E_{0}}}
&+ \sigma^{(\varphi\varphi)}\left\{ C_{l}^{E_{0}\partial^2 E_{0}}
+ \frac{\theta_c^2}{4}\int dl' l' e^{-\frac{1}{4}(l^2+{l'}^2)\theta_c^2} \right
.\nonumber \\
&\times \left. \left[I_{0} \left(\frac{\theta_{c}^2}{2}l l' \right) + I_{4}
\left(\frac{\theta_{c}^2}{2}l l' \right)\right]  C_{l'}^{\partial E_{0}\partial
E_{0}} \right\} \, .
\end{align}
We plot this average power spectrum in Fig.~{\ref{fig:EE_powspect}.  Again, the
effect of $\varphi$ fluctuations is to reduce the amplitude of the
oscillatory component at small angular scales.

\subsection{The $BB$ Power Spectrum}

Using Eq.~(\ref{eqn:b}), and once again dropping the correlators involving an
odd number of fields and those that vanish due to parity, we find that the
expansion for the two point correlation function is
\begin{align}
\langle B(&\vect{l})B(\vect{l'}) \rangle =
\<{ B_{0}(\vect{l})B_{0}(\vect{l'}) } \nonumber
\\
&+ \frac{1}{2}\<{
B_{0}(\vect{l})\left[\pphi^2B_{0}\convc\varphi\conv\varphi\right](\vect{l'}) }
\nonumber
\\
&+ \frac{1}{2}\<{
\left[\pphi^2B_{0}\convc\varphi\conv\varphi\right](\vect{l})B_{0}(\vect{l'}) }
\nonumber
\\
&+ \<{ \left[\pphi B_{0}\convc\varphi\right](\vect{l})\left[\pphi
B_{0}\convc\varphi\right](\vect{l'}) }
\nonumber \\
&+ \<{ \left[\pphi E_{0}\convs\varphi\right](\vect{l})\left[\pphi
E_{0}\convs\varphi\right](\vect{l'}) }
\end{align}
to $O(\varphi^2)$.

Evaluating the convolutions and Wick expanding these correlators
leads to the expressions
\begin{align}
\langle & B_{0}(\vect{l})\left[\pphi^2  B_{0}\convc\varphi\conv
\varphi\right](\vect{l'})  \rangle = \langle \left[\pphi^2
B_{0}\convc\varphi\conv \varphi\right](\vect{l'}) B_{0}(\vect{l})  \rangle
\nonumber \\
& =(2\pi)^2 \delta^2(\vect{l}+\vect{l'}) \left[\sigma^{(\varphi\varphi)}C_{l}^{
B_{0}\partial^2 B_{0}} \right] \, ,
\end{align}
\begin{align}
\langle &\left[\pphi B_{0}\convc\varphi\right](\vect{l})\left[\pphi
B_{0}\convc\varphi\right](\vect{l'}) \rangle \nonumber \\
& =(2\pi)^2 \delta^2(\vect{l}+\vect{l'})\int \frac{d^2 \vect{l}_1}{(2\pi)^2}
 \cos^2{(2\phi_{l_1})}C_{l_1}^{\partial B_{0}\partial B_{0}}
C_{|\vect{l}-\vect{l}_1|}^{\varphi\varphi}  \, ,
\end{align}
\begin{align}
\langle &\left[\pphi E_{0}\convs\varphi\right](\vect{l})\left[\pphi
E_{0}\convs\varphi\right](\vect{l'}) \rangle \nonumber \\
& =(2\pi)^2 \delta^2(\vect{l}+\vect{l'})\int \frac{d^2 \vect{l}_1}{(2\pi)^2}
\left\{ \sin^2{(2\phi_{l_1})}C_{l_1}^{\partial E_{0}\partial E_{0}}
C_{|\vect{l}-\vect{l}_1|}^{\varphi\varphi} \right. \nonumber \\
&+ \sin{(2\phi_{l_1})}\left[\frac{l_1^2 \sin(2\phi_{l_1})-2ll_1 \sin{(\phi_{l_1})}
}{
|\vect{l}-\vect{l}_1|^2} \right] \nonumber \\
& \hspace{1.6cm} \left. \times C_{l_1}^{\partial
E_{0}\varphi}C_{|\vect{l}-\vect{l}_1|}^{\partial E_{0}\varphi} \right\} \, .
\end{align}
Any $\varphi$-$B_{0}$ cross correlations must vanish due to parity.

\begin{figure}
\centerline{\psfig{file=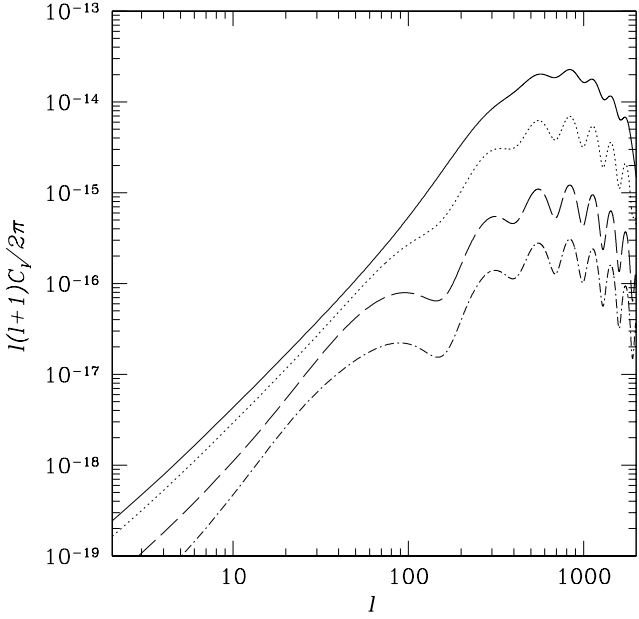,width=3.67in,angle=0}}
\caption{The $B B$ power spectra for the Gaussian correlation function for
$\varphi$ with $\sigma^{(\varphi \varphi)}=9\times 10^{-4}$ and
$\theta_{c}=1^{\circ}$ (solid), $\theta_{c}=2^{\circ}$ (dotted),
$\theta_{c}=5^{\circ}$ (dashed), and $\theta_{c}=10^{\circ}$ (dot-dashed).}
\label{fig:BB_powspect}
\end{figure}

Collecting all terms we
find that the expression for the average $B$-mode power spectrum is
\begin{align}
&C_{l}^{BB} = C_{l}^{B_{0}B_{0}} + \sigma^{(\varphi\varphi)}
C_{l}^{B_{0}\partial^2 B_{0}} \nonumber \\
&+ \int \frac{d^2 \vect{l'}}{(2\pi)^2} \left\{
\left[ \cos^2{(2\phi_{l'})}C_{l'}^{\partial B_{0}\partial B_{0}} \right.
\right. \nonumber \\
&+ \left. \left. \sin^2{(2\phi_{l'})}C_{l'}^{\partial E_{0}\partial E_{0}}
\right] C_{|\vect{l}-\vect{l'}|}^{\varphi\varphi} \right.
\nonumber \\
&+ \sin{(2\phi_{l'})}\left[\frac{l'^2 \sin(2\phi_{l'})-2ll' \sin{(\phi_{l'})}
}{
|\vect{l}-\vect{l'}|^2} \right] \left. C_{l'}^{\partial
B_{0}\varphi}C_{|\vect{l}-\vect{l'}|}^{\partial B_{0}\varphi} \right\} \,  ,
\end{align}
and if $E_{0}$ and $\varphi$ are uncorrelated the expression simplifies to
\begin{align}
C_{l}^{BB} = C_{l}^{B_{0}B_{0}} + \sigma^{(\varphi\varphi)}
C_{l}^{B_{0}\partial^2 B_{0}} \hspace{-0.72cm} &
\nonumber \\
+ \int \frac{d^2 \vect{l'}}{(2\pi)^2}
 [ & \cos^2{(2\phi_{l'})}C_{l'}^{\partial B_{0}\partial B_{0}}
\nonumber \\
+ & \sin^2{(2\phi_{l'})}C_{l'}^{\partial E_{0}\partial E_{0}} ]
C_{|\vect{l}-\vect{l'}|}^{\varphi\varphi}
\, .
\end{align}

If the intrinsic $B$-modes due to gravitational waves are negligible then
for our $\varphi$ power spectrum, Eq.~(\ref{eqn:phipower}), the average
$B$-mode power spectrum is
\begin{align}
C_{l}^{{B B}} = \sigma^{(\varphi\varphi)}& \left\{ \frac{\theta_c^2}{4}\int dl'
l' e^{-\frac{1}{4}(l^2+{l'}^2)\theta_c^2} \right .\nonumber \\
&\times \left[I_{0} \left(\frac{\theta_{c}^2}{2}l l' \right) - I_{4}
\left(\frac{\theta_{c}^2}{2}l l' \right)\right] \left. C_{l'}^{\partial
E_{0}\partial E_{0}} \right\} \, .
\end{align}
This expression shows that $E$ modes modulated by $\varphi$ fluctuations can
induce $B$ modes.  We note here that this effect is more general than the
specific application of $\alpha$ variation we focus on here, and is a generic
feature of any fluctuations that modulate the power spectrum but have
negligible effects on the evolution of the dominant density perturbations. In
Fig.~{\ref{fig:BB_powspect}} we plot the induced $B$-mode power spectra due to
$\varphi$ fluctuations.  The induced $B$-mode power spectrum inherits the
oscillatory features of the unperturbed $E_{0}$-mode power spectrum as long as
correlation angle is larger than the horizon size at recombination, while the
amplitude of the power spectrum decreases as the correlation angle increases.

\section{Bispectra}
\label{sec:Bispect}

We have shown in the previous Section that modulation of the temperature and
polarization power spectra by spatial variations of $\alpha$ alters the mean
power spectra of CMB anisotropies.  In this Section and the one that follows we
show how this modulation introduces higher order correlations into the CMB
temperature field.

The temperature bispectrum is defined in terms of the connected piece (the
terms remaining after the Gaussian piece is subtracted out) of the three
point correlation function in Fourier space as
\begin{align}
\langle \Theta(\vect{l}_1)& \Theta(\vect{l}_2)\Theta(\vect{l}_3) \rangle_c
\nonumber \\
&= (2\pi)^2\delta^2(\vect{l}_1+\vect{l}_2+\vect{l}_3
)B^{{\Theta\Theta\Theta}}(\vect{l}_1,\vect{l}_2,\vect{l}_3) \, .
\end{align}

This expression must be invariant under the exchange of any two
fields, or equivalently, $\vect{l}$ vectors.  We insert Eq.~(\ref{eqn:theta})
into
the three-point correlation function, and after some straightforward algebra we
find that
\begin{align}
B^{{\Theta\Theta\Theta}}(\vect{l}_1,\vect{l}_2,\vect{l}_3)
= &\sum_{\substack{i,j=1 \\ i < j}}^3 {\rm
B}(\vect{l}_{i},\vect{l}_{j},\vect{l}_{6-i-j}) \, ,
\label{eqn:bi1}
\end{align}
where the sums runs over the distinct permutations of $\{1,2,3\}$ and, to
leading order,
\begin{align}
{\rm B}(\vect{l}_{i},\vect{l}_{j},\vect{l}_{k})
&= \, C_{l_{i}}^{ \Theta_{0}\partial\Theta_{0} }
C_{l_{j}}^{{\Theta_{0}\varphi}} +  C_{l_{i}}^{\Theta_{0}\varphi}
C_{l_{j}}^{{\Theta_{0}\partial\Theta_{0}}}
\, .
\label{eqn:bi2}
\end{align}

Thus the temperature bispectrum vanishes unless $\varphi$ and $\Theta_{0}$ are
correlated. In the models we consider in this work, $\varphi$ and $\Theta_{0}$
are not expected to be highly correlated.  However, if a model of spatial
variations of $\alpha$ predicts that variations of $\varphi$ and $\Theta_{0}$
are strongly correlated the bispectrum may be a strong signature of such a
model as, unlike the power spectrum or the trispectrum (discussed below), it is
first order in $\varphi$.  We note that expressions analogous to
Eqs.~(\ref{eqn:bi1}) and (\ref{eqn:bi2}) will hold for the polarization and
cross-bispectra as well.

\section{Trispectra}
\label{sec:Trispect}

In analogy with the bispectrum, the temperature trispectrum is defined in terms
of the connected piece of the four point correlation function in Fourier space
as
\begin{align}
\langle \Theta(\vect{l}_1)&
\Theta(\vect{l}_2)\Theta(\vect{l}_3)\Theta(\vect{l}_4) \rangle_c \nonumber \\
&=
(2\pi)^2\delta^2(\vect{l}_1 +\vect{l}_2+\vect{l}_3
+\vect{l}_4)T^{{\Theta\Theta\Theta\Theta}}
(\vect{l}_1,\vect{l}_2,\vect{l}_3,\vect{l}_4)\, .
\end{align}
This expression must also be invariant under the exchange of any two
fields, or equivalently, $\vect{l}$ vectors.  We insert Eq.~(\ref{eqn:theta})
into
the four-point correlation function, and after some straightforward algebra we
find that
\begin{align}
T^{{\Theta\Theta\Theta\Theta}}(\vect{l}_1,\vect{l}_2,\vect{l}_3,\vect{l}_4)
= &\sum_{\substack{i,j=1 \\ i < j}}^4 \sum_{\substack{k,l=1 (\neq i,j) \\ k <
l}}^4{\rm T}_{\rm A} (\vect{l}_{i},\vect{l}_{j},\vect{l}_{k},\vect{l}_{l})
\nonumber \\
&+ \sum_{\substack{i,j,k=1 \\ i < j < k}}^4 \sum_{\substack{l=1 \\ l \neq
i,j,k}}^4 {\rm T}_{\rm B} (\vect{l}_{i},\vect{l}_{j},\vect{l}_{k},\vect{l}_{l})
\, ,
\label{eqn:tri1}
\end{align}
where the sums runs over the distinct permutations of $\{1,2,3,4\}$,
\begin{align}
{\rm T_{\rm A}}&(\vect{l}_{i},\vect{l}_{j},\vect{l}_{k},\vect{l}_{l})
= C_{l_{i}}^{{\Theta_{0}\partial\Theta_{0}}}
C_{l_{j}}^{{\Theta_{0}\partial\Theta_{0}}} \left( C_{|\vect{l}_j +
\vect{l}_k|}^{\varphi\varphi}+
C_{|\vect{l}_j + \vect{l}_l|}^{\varphi\varphi} \right) \nonumber \\
&+
\left( C_{l_{i}}^{{\Theta_{0}\partial\Theta_{0}}}
C_{l_{j}}^{{\Theta_{0}\varphi}} + C_{l_{i}}^{{\Theta_{0}\varphi}}
C_{l_{j}}^{{\Theta_{0}\partial\Theta_{0}}} \right) \left( C_{|\vect{l}_j +
\vect{l}_k|}^{\partial\Theta_{0}\varphi}+
C_{|\vect{l}_j + \vect{l}_l|}^{\partial\Theta_{0}\varphi} \right) \nonumber \\
&+
C_{l_{i}}^{{\Theta_{0}\varphi}} C_{l_{j}}^{{\Theta_{0}\varphi}} \left(
C_{|\vect{l}_j + \vect{l}_k|}^{\partial\Theta_{0}\partial\Theta_{0}}+
C_{|\vect{l}_j + \vect{l}_l|}^{\partial\Theta_{0}\partial\Theta_{0}} \right)
\, ,
\end{align}
and
\begin{align}
{\rm T_{\rm B}}&(\vect{l}_{i},\vect{l}_{j},\vect{l}_{k},\vect{l}_{l})
= 2C_{l_{i}}^{{\Theta_{0}\partial^2\Theta_{0}}} C_{l_{j}}^{{\Theta_{0}\varphi}}
  C_{l_{k}}^{{\Theta_{0}\varphi}}+ \nonumber \\
&+ 2C_{l_{i}}^{{\Theta_{0}\varphi}} (C_{l_{j}}^{{\Theta_{0}\varphi}}
C_{l_{k}}^{{\Theta_{0}\partial^2\Theta_{0}}} +
C_{l_{j}}^{{\Theta_{0}\partial^2\Theta_{0}}}   C_{l_{k}}^{{\Theta_{0}\varphi}})
\, .
\label{eqn:tri3}
\end{align}

If $\Theta_{0}$ and $\varphi$ have no cross correlation, the second term
vanishes and the first term simplifies to
\begin{align}
{\rm T_{A}}(\vect{l}_{i},\vect{l}_{j},\vect{l}_{k},\vect{l}_{l})
= \, C_{l_{i}}^{{\Theta_{0}\partial\Theta_{0}}} &
C_{l_{j}}^{{\Theta_{0}\partial\Theta_{0}}} \left( C_{|\vect{l}_j +
\vect{l}_k|}^{\varphi\varphi}+
C_{|\vect{l}_j + \vect{l}_l|}^{\varphi\varphi} \right) \, .
\label{eqn:simp_tri}
\end{align}
Expressions similar to Eqs.~(\ref{eqn:tri1})-(\ref{eqn:tri3}) will hold for the
polarization and cross-trispectra as well.

\subsection{The Kurtosis}

The trispectrum is a nontrivial function of 6 variables and extracting
the full trispectrum from CMB data will be a challenging experimental endeavor.
It is therefore worthwhile to examine simpler statistical signatures of
non-Gaussianity in the CMB.  If the trispectrum is nonvanishing the probability
distribution function of $\Theta(\vhat{n})$ on the sky will no longer be
precisely Gaussian. As in the
case of weak lensing, the deviation from Gaussianity can be parametrized by the
kurtosis \cite{Kesden02}, which is a measure of how flattened out or peaked the
distribution is relative to a Gaussian distribution.

The kurtosis of a non-Gaussian random field may be written in terms
of the trispectrum as
\begin{align}
K(\theta)&=\frac{1}{\sigma^4(\theta)} \int
\frac{d^2\vect{l}_1}{(2\pi)^2}\frac{d^2\vect{l}_2}{(2\pi)^2}
\frac{d^2\vect{l}_3}{(2\pi)^2}\frac{d^2\vect{l}_4}{(2\pi)^2} \nonumber \\
&\times \left\{
(2\pi)^2\delta^2(\vect{l}_1 +\vect{l}_2+\vect{l}_3+ \vect{l}_4)
T^{{\Theta\Theta\Theta\Theta}}(\vect{l}_1,\vect{l}_2,\vect{l}_3,\vect{l}_4)
\right. \nonumber \\
&\times W(l_{1}\theta)W(l_{2}\theta)W(l_{3}\theta)W(l_{4}\theta) \big{\}} \, ,
\end{align}
where $W(l\theta)$ is a smoothing function with smoothing scale $\theta$, and
\begin{align}
\sigma^2(\theta)=\int
\frac{d^2\vect{l}}{(2\pi)^2}C_{l}^{\Theta\Theta}W^2(l\theta)
\end{align}
is the smoothed variance.

If we adopt a Gaussian smoothing function
\begin{align}
W(l\theta)=e^{-\frac{1}{2}\sigma_b^2l^2}
\end{align}
with $\sigma_{b}=\theta/(\sqrt{8 {\rm ln}2})$ and insert the expression
for the trispectrum from Eq.~(\ref{eqn:simp_tri}) we find after some algebra
that
\begin{align}
K(\theta)=&\frac{3}{2\pi^3 \sigma^4(\theta)}\int l_1 dl_1 l_2 dl_2 l_3 dl_3
\Big{\{} C_{l_1}^{\Theta\partial\Theta}
C_{l_2}^{\varphi\varphi}C_{l_3}^{\Theta\partial\Theta}  \nonumber \\
&\times \left[ I_{0}(\sigma_b^2 l_1 l_2) I_{0}(\sigma_b^2 l_2 l_3)
e^{-\sigma_b^2(l_1^2+l_2^2+l_3^2)}\right] \Big{\}} \, .
\end{align}

\begin{figure}
\centerline{\psfig{file=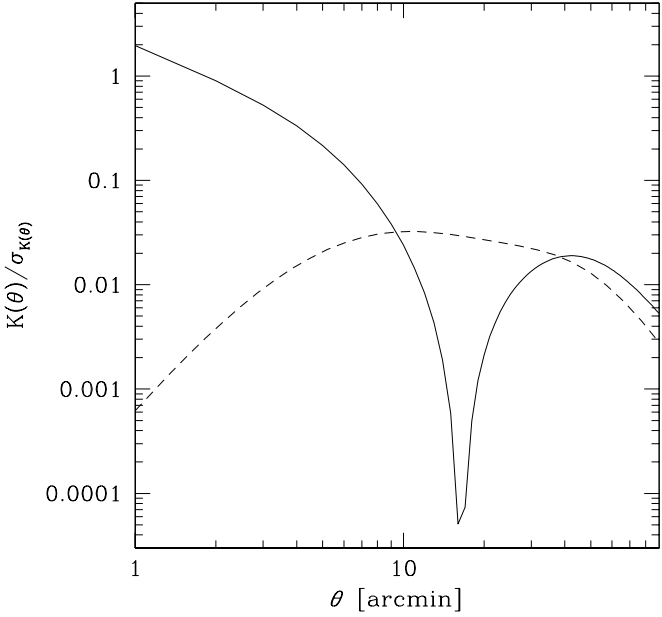,width=3.67in,angle=0}}
\caption{The ratio of the kurtosis to the variance in the estimator of the
kurtosis due to Gaussian fluctuations for spatial $\alpha$ fluctuations with
$\sigma^{(\varphi \varphi)}=9\times 10^{-4}$ and $\theta_{c}=1^{\circ}$ (solid)
and the same ratio for weak lensing (dashed).  The kurtosis is undetectable for
weak lensing at any angular resolution, while for spatial $\alpha$ fluctuations
of this amplitude the kurtosis may in principle be detected by a high
resolution no-noise experiment.}
\label{fig:kurt_sn}
\end{figure}

It was shown in Ref.~\cite{Kesden02} that the sample variance of the kurtosis
of a Gaussian random field smoothed over an angular scale $\theta$ for a
full-sky experiment is
\begin{equation}
\sigma^2_{K(\theta)}=\frac{3}{2}\theta^2 \, .
\end{equation}
In Fig.~{\ref{fig:kurt_sn}} we plot the ratio of the kurtosis to the standard
deviation of the kurtosis as a function of smoothing scale.  Despite the
improved variance at high resolution the kurtosis due to weak lensing cannot be
detected because in the limit of infinite resolution the kurtosis of weak
lensing vanishes more quickly than the variance.  This can be understood by
noting that weak lensing is power conserving because it maps the temperature at
one point on the sky to another, and so with infinite resolution the
probability distribution function is Gaussian.  In contrast, the kurtosis due
to spatial $\alpha$ fluctuations approaches a constant value in the limit of
infinite resolution, and so for low $\theta$ the signal-to-noise increases like
$\theta^{-1}$.  This occurs because $\alpha$ fluctuations do not lead to a pure
remapping of of the temperature pattern (they do not conserve power), and thus
in the limit of infinite resolution modulate the variance of the (Gaussian)
temperature probability distribution function from one patch of the sky to
another.  The resulting mean probability distribution is no longer Gaussian.
Note that the dip in the kurtosis at $\theta \approx 15$ arcminutes occurs
because when smoothed over that scale modulation of the CMB due to spatial
$\alpha$ fluctuations becomes approximately power conserving.  An ideal
noise-free experiment at an angular resolution of 1 arcminute could detect the
kurtosis due to $\alpha$ fluctuations at the level of $\sigma^{(\varphi
\varphi)}=9\times 10^{-4}$, while higher resolution experiments could observe
lower amplitude $\alpha$ fluctuations.

\subsection{A Discriminating Filter}

As discussed above, weak lensing of the CMB by matter along the line of sight
must induce a contribution to the trispectrum
\begin{align}
L^{{\Theta\Theta\Theta\Theta}}(\vect{l}_1,\vect{l}_2,\vect{l}_3,\vect{l}_4)
= &\sum_{\substack{i,j=1 \\ i < j}}^4 \sum_{\substack{k,l=1 (\neq i,j) \\ k <
l}}^4{\rm L}(\vect{l}_{i},\vect{l}_{j},\vect{l}_{k},\vect{l}_{l}) \, ,
\end{align}
where
\begin{align}
{\rm L}(\vect{l}_{i},& \vect{l}_{j},\vect{l}_{k},\vect{l}_{l})
= -\, C_{l_{k}}^{{\Theta_{0}\Theta_{0}}} C_{l_{l}}^{{\Theta_{0}\Theta_{0}}}
\times \nonumber \\
&\times \left[ C_{|\vect{l}_i + \vect{l}_k|}^{\phi\phi}[(\vect{l}_i +
\vect{l}_k)\cdot\vect{l}_k][(\vect{l}_i + \vect{l}_k)\cdot\vect{l}_l] \right.
\nonumber \\
&+ \left. C_{|\vect{l}_j + \vect{l}_k|}^{\phi\phi}[(\vect{l}_j +
\vect{l}_k)\cdot\vect{l}_k][(\vect{l}_j + \vect{l}_k)\cdot\vect{l}_l \right] \,
{}.
\end{align}
We emphasize that in this expression $C_l^{\phi\phi}$ is the power spectrum of
the projected lensing potential and not the power spectrum of $\alpha$
fluctuations $C_l^{\varphi\varphi}$.

Since this contribution to the trispectrum must be in the CMB, we now derive
the a filter to distinguish between the weak-lensing and $\alpha$-fluctuation
trispectra.

For compactness we introduce the notation
\begin{align}
X_{ijk} \equiv
X^{{\Theta\Theta\Theta\Theta}}(\vect{l}_i,\vect{l}_j,
\vect{l}_k,-\vect{l}_i-\vect{l}_j-\vect{l}_k) \, ,
\end{align}
and
\begin{align}
\sum_{ijk} \equiv
\int \frac{d^2\vect{l}_i}{(2\pi)^2}\frac{d^2\vect{l}_j}{(2\pi)^2}
\frac{d^2\vect{l}_k}{(2\pi)^2} \, .
\end{align}

The signal for detection of the $\alpha$-fluctuation trispectrum can be
in general written as a windowed integral over trispectrum configurations
(quadrilaterals in $\vect{l}$-space),
\begin{align}
S=\sum_{ijk} W_{ijk}T_{ijk} \, ,
\end{align}
where $W_{ijk}$ is the to-be-determined window function.

To determine the least-square quartic discriminator we treat the trispectrum due to
weak gravitational lensing as a source of noise that should be minimized
and write
\begin{align}
N_{L}=\sum_{ijk} W_{ijk}L_{ijk} \, .
\end{align}

In addition we at minimum also have the Gaussian noise due to cosmic
variance
\begin{align}
N_{G}^2=\sum_{ijk} W^2_{ijk}G_{ijk} \, ,
\end{align}
where
\begin{align}
G_{ijk}=\frac{4!}{4\pi
f_{sky}}C_{l_i}^{\widehat{{\Theta\Theta}}}
C_{l_j}^{\widehat{{\Theta\Theta}}}C_{l_k}^{\widehat{{\Theta\Theta}}}
C_{|\vect{l}_i+\vect{l}_j+\vect{l}_k|}^{\widehat{{\Theta\Theta}}} \, ,
\end{align}
and
\begin{align}
C_{l}^{\widehat{{\Theta\Theta}}}=C_l^{\Theta\Theta} + \frac{4 \pi f_{sky}
s^2}{t_{exp} T_{CMB}^2}e^{\sigma_b^2 l^2}
\end{align}
is the sum of the actual CMB power spectrum and the noise power spectrum
introduced by an experiment that observes a fraction $f_{sky}$ of the sky with
beam width $\sigma_b$ for a time $t_{exp}$ with detectors of
noise-equivalent-temperature $s$.

The total noise is then just
\begin{align}
N^2=N_{G}^2+N_{L}^2 \, .
\end{align}
We want to solve for the $W_{ijk}$ that maximizes the signal to noise
ratio ${S}/{N}$.   To do this we set
\begin{align}
\frac{\delta}{\delta W_{ijk}}\left(\frac{S^2}{N^2} \right) = 0 \, .
\end{align}
After some straightforward algebra, and making use of the fact that
${S^2}/{N^2}$ is invariant under renormalization of $W_{ijk}$, we find that the
signal-to-noise ratio is extremized only if $W_{ijk}$ takes the form
\begin{align}
W_{ijk}=\frac{T_{ijk}- \lambda L_{ijk}}{G_{ijk}} \, .
\end{align}
Thus we have
\begin{align}
\frac{S^2}{N^2}=\frac{{\cal T}^2 - 2\lambda{\cal T}{\cal X} + \lambda^2 {\cal
X}^2}{{\cal T} + {\cal X}^2 - (2\lambda{\cal X}-\lambda^2{\cal L})(1+{\cal L})}
\, ,
\end{align}
where
\begin{align}
{\cal T}=\sum_{ijk} \frac{T^2_{ijk}}{G_{ijk}} \, ,
\end{align}
\begin{align}
{\cal X}=\sum_{ijk} \frac{L_{ijk}T_{ijk}}{G_{ijk}} \, ,
\end{align}
\begin{align}
{\cal L}=\sum_{ijk} \frac{L^2_{ijk}}{G_{ijk}} \, ,
\end{align}
and the optimum weighting is
\begin{equation}
\lambda=\frac{{\cal X}}{1+{\cal L}}\, .
\end{equation}

In the absence of lensing this expression reduces to the conventional
signal-to-noise measure
\begin{align}
\frac{S^2}{N^2}={\cal T}=\sum_{ijk} \frac{T^2_{ijk}}{G_{ijk}} \, .
\end{align}

\begin{figure}
\centerline{\psfig{file=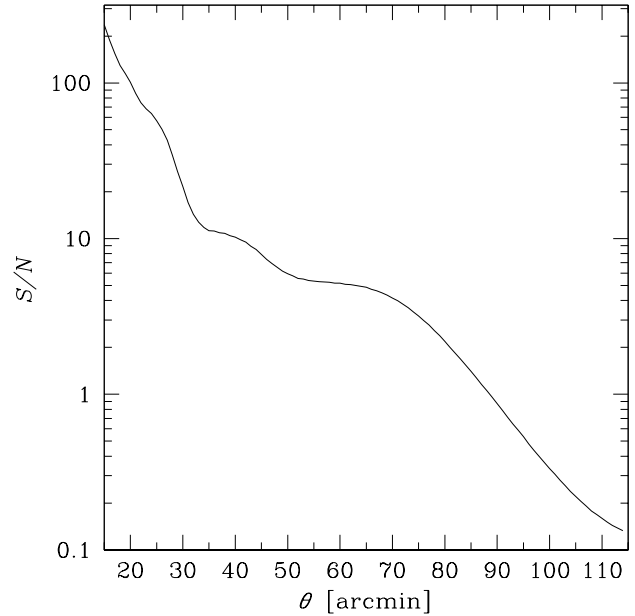,width=3.67in,angle=0}}
\caption{The signal-to-noise ratio of the $\alpha$-fluctuation
trispectrum that would be observed by an experiment with
$s=12.42 ~\mu {\rm K} \sqrt{{\rm sec}}$ observing for 1 year (the
effective parameters of the Planck satellite) with resolution
$\theta$ if weak lensing were not present for
$\sigma_{(\varphi\varphi)}=9\times 10^{-4}$. Since in practice
we know weak lensing must be present in the CMB this curve
serves as an upper bound on the observable signal-to-noise
ratio.}
\label{fig:SN_alpha}
\end{figure}

In Fig.~\ref{fig:SN_alpha} we show the maximum signal-to-noise ratio for
detection of the $\alpha$-fluctuation trispectrum that would be detected were
weak lensing not present.  For a given weak lensing power spectrum
$C_{l}^{\phi\phi}$ the actual signal-to-noise will be less than the bound
shown.

\section{Theoretical Models of Variable {\large $\alpha$}}
\label{sec:ParticleTheory}

Recent theoretical work
\cite{Uzan03,Maretal03,Battye01,Sandvik02,Youm02,
Barrow02d,Huang03,Dvali02,Bek02,Chiba02} has considered
the ingredients required of field-theory models for variable
fine-structure parameter in order to explain a small time
variation of $\alpha$.  Here we briefly amend those discussions
to consider the field-theory requirements for spatial variations
of the sort we consider in this paper.

The simplest way to introduce spatial variation of $\alpha$ is
to couple the photon to a scalar field $\phi(x)$ through a term
(see, e.g., Ref. \cite{Bek02} and references therein)
\begin{equation}
     {\cal L}_{\phi\gamma\gamma} = -\frac{\lambda}{4}
g\left(\frac{\phi}{f_\phi} \right)
     F^{\mu\nu} F_{\mu\nu},
\label{eqn:interaction}
\end{equation}
in the Lagrangian that appears in addition to the usual
electromagnetic Lagrangian, $F^{\mu\nu} F_{\mu\nu}/4$.  Here,
$f_\phi$ is a constant with dimensions of mass, $\lambda$ is an overall
coupling constant, and $g(x)=\sum_{n=1}^{\infty} c_{n}x^n$ is a dimensionless
function of the ratio $\phi/f_\phi$ normalized so that $g(1)=1$.  This
interaction  then leads to a fractional change in
$\alpha$,
\begin{equation}
     \frac{\delta\alpha}{\alpha} = -\lambda g \left(\frac{\phi}{f_\phi}\right) = \varphi \, .
\end{equation}
The best present constraint to $g$ comes from the evolution
of globular-cluster stars; this requires $[\lambda(dg/d\phi)]^{-1} \gtrsim
1.6\times10^{10}$~GeV \cite{Hagetal02}.
If we consider fractional fluctuations in $\alpha$ of $\lesssim 10^{-2}$, then
$\lambda \sim 10^{-2}$, $\delta\phi \lesssim f_{\phi}$, and $\sigma^{(\varphi \varphi)} \sim \lambda^2$.

Cosmological mechanisms that might induce spatial variations in
$\phi$ are analogous to those that have been considered, e.g.,
for spatial variations in the axion field \cite{axions}.
These variations in $\phi$ can arise either during or after
inflation.  If $\phi$ is a spectator field during inflation,
then fluctuations in $\phi$ can be induced quantum mechanically
during inflation resulting in a nearly scale-invariant spectrum
for $\phi$ fluctuations.  Alternatively, if $\phi$ is a
pseudo-Nambu-Goldstone field in a model with an approximate global symmetry, it
could fall during the symmetry-breaking phase transition to
different random points on the vacuum manifold (which is periodic in $\phi$
with period $2f_{\phi}$) in different
causally-disconnected regions.  In this case, the scalar-photon
interaction will most generally be an arbitrary Fourier series in $\phi$.  A
simple interaction of this form, containing only the first harmonic and (as
motivated below) constraining $g(\phi/f_{\phi})$ to be an even function, is
\begin{equation}
     {\cal L}_{\phi\gamma\gamma} = -\frac{\lambda}{8}
\left[ 1 -     \cos\left( \frac{\pi \phi}{f_\phi} \right) \right]F^{\mu\nu}
     F_{\mu\nu} \, .
\end{equation}
For such an interaction $\phi$ will be uncorrelated on scales larger than the
horizon, and
$\delta\alpha/\alpha$ is fixed by the magnitude of the explicit
symmetry breaking term $\lambda $, rather than by $f_\phi$.  In
either case (inflation or spontaneous symmetry breaking), the gradient term in the
scalar-field Lagrangian
will tend to align the scalar field within causally-connected
regions of the Universe.  Thus, the value of $\alpha$ at the
surface of last scatter should be constant within
square-degree patches, but will vary from one
square-degree patch to another, as we have assumed throughout
this paper.

Now consider constraints to the potential-energy density
$V(\phi)$ for the scalar field.  If it is quadratic,
$V(\phi)=m_\phi^2 \phi^2/2$, then (neglecting for the moment the interaction
with photons) the equation of motion for $\phi$ is
\begin{align}
\ddot{\phi} + 3H\dot{\phi} + m_{\phi}^2 \phi = 0 \,
\end{align}
and has the solution
\begin{align}
\phi(t)=\phi_{0}\frac{\sin(m_{\phi}t)}{m_{\phi}t}
\label{eqn:phi_soln}
\end{align}
in the matter dominated era.  At early times when $H \gg m_{\phi}$ the value of
$\phi$ is frozen at $\phi \approx \phi_{0}$, while for $H \lesssim m_{\phi}$
the field in each horizon will oscillate about its minimum with an amplitude
decaying like $1/t$.  We thus require $m_\phi \lesssim H_{\rm rec}
\simeq 10^{-28}$~eV so that the $\alpha$ fluctuations are frozen in at
recombination.  On the other hand, observations of the
Lyman-$\alpha$ forest at redshifts $z\sim4$ show no evidence of
spatial fluctuations in $\alpha$ larger than one part in
$10^4$.  Thus, if we consider
$(\delta\alpha/\alpha)\gtrsim10^{-4}$ at the surface of last
scatter, then we must require that $\lambda g|_{z \approx 4} \lesssim 10^{-4}$.
If the leading order term in $g$ is quadratic in $\phi$ (as we argue below)
this leads to the constraint $ m_{\phi} \gtrsim \sqrt{\lambda/0.01} \times
10^{-31}$~eV.  Finally, laboratory experiments constrain the value
$\dot{\alpha}/\alpha$ today to be $ \lesssim 5 \times 10^{-15}~\rm{yr}^{-1}$
\cite{Soretal01}.  To satisfy this requirement we require that $g \propto
\phi^2$ to leading order at small $\phi$ so that $\dot{\alpha}/\alpha \propto
t^{-2}$.  This leads to the constraint $ m_{\phi} \gtrsim 2 \times
(\lambda/0.01)\times 10^{-30}$~eV if the oscillatory factor is unity.  We note
that if $g \propto \phi^n$ with $n > 2$ for small $\phi$ this constraint can be
relaxed, and that late-time $\Lambda$ domination does not significantly alter
the limit.  These constraints will insure that the model conforms to upper
limits to $\alpha$ variations in the low-redshift Universe, but we caution that
they are constraints based on the root-mean-squared value of the scalar field.
If we happen to live in a region where the amplitude of $\phi$ is randomly
small (large) the constraints based on local observations will be weaker
(stronger) than discussed above.  Also, the nonlinear evolution of the mass distribution may modify constraints based on recent terrestrial phenomena \cite{Mota03}.

We must also be sure that fluctuations in the scalar field do
not lead to density perturbations that exceed those of amplitude
$10^{-5}$.  This constraint requires that the gradient-energy
density, $k^2 (\Delta\phi)^2 \lesssim 10^{-5}\,\rho_m(z_{\rm rec})$,
where $k\simeq H_{\rm rec}$ is the largest wavenumber for which
perturbations are significant at the redshift $z_{\rm
rec}\simeq1100$ of decoupling, and $H_{\rm rec}=\Omega_m^{1/2} H_0
(1+z_{\rm rec})^{3/2}$ is the Hubble parameter at decoupling.
However, from the Friedmann equation, $\rho_m(z_{\rm rec}) \sim
H_{\rm rec}^2 m_{\rm Pl}^2$, where $m_{\rm Pl}\simeq
10^{19}$~GeV is the Planck mass, so we find a constraint
$\Delta\phi \lesssim f_{\phi} \lesssim 3\times10^{16}$~GeV.

Now lets examine the effect of the scalar-photon Lagrangian on the scalar field
dynamics.  If we include the interaction in the Lagrangian it is easy to verify
the equation of motion for the scalar field is modified to include a forcing
term
\begin{align}
\ddot{\phi} + 3H\dot{\phi} + m_{\phi}^2 \phi +
\frac{\lambda}{4}\frac{dg}{d\phi}\left<F_{\mu\nu}F^{\mu\nu}\right>= 0 \, ,
\end{align}
where
\begin{align}
\left<F_{\mu\nu}F^{\mu\nu}\right>=\frac{1}{8\pi}\left(\left< E^2 \right> -\left< B^2 \right>
\right) \,
\end{align}
and $\left< E^2 \right>$ and $\left< B^2 \right> $ are the
spatially averaged squared electric and magnetic fields
respectively.  A bath of thermal photons has $\left< E^2 \right>
= \left< B^2 \right> $ and so will not contribute to the forcing
term.  However, nonrelativistic matter is a source for
electromagnetic fields with $\left< E^2 \right> \neq \left< B^2
\right> $, and so the forcing term should be proportional to the
density of nonrelativistic matter $\rho_m$.  If, as we discussed
above, $g(\phi/f_{\phi})$ is dominated by its quadratic term
then in the matter dominated era the equation of motion reads
\begin{align}
\ddot{\phi} + \frac{2}{t}\dot{\phi} + (m_{\phi}^2 + \frac{\eta}{t^2}) \phi = 0
\, ,
\end{align}
where $\eta=(2 \lambda \xi \rho_{m_0} t_0^2)/f_{\phi}^2$, and $\xi$ is
the fraction of matter density $\rho_{m}$ due to electromagnetic
energy.\footnote{Naively, one might guess that $\sim$nG
intergalactic magnetic fields would suggest $\xi\sim10^{-11}$.
However, the quantity of interest here is $\langle B^2 \rangle$,
rather than $\langle B \rangle^2$, as well as $\langle E^2
\rangle$.  Thus, strong microscopic magnetic fields in the vicinity of
electrons and protons may contribute $\xi_b\sim10^{-5}$
\cite{Bek02}.}  Thus, if $\eta$ is small the effect of the
forcing term
is to introduce a small time-dependent mass term into the
equation of motion.  The solution will qualitatively behave like
that of Eq.~(\ref{eqn:phi_soln}), except that the time that
field begins to oscillate may shift by a small $\eta$-dependent
factor, and the power law of decay will shift from $-1$ to
$-1+\eta$.  Generically $\eta$ need not be small, but for
$f_{\phi} \gtrsim 5 \times 10^{15}$~GeV and $\lambda \approx 0.01$ we find  $|\eta| \lesssim 0.1$~.\footnote{This assumes a baryon fraction $\Omega_b/\Omega_m=1/6$, that magnetic energy contributes $|\xi_{b}| \approx 2 \times 10^{-5}$ \cite{Bek02}, and that the dark matter has negligible electromagnetic energy so that $\xi = (\Omega_b/\Omega_m)\xi_b$.} This brings us uncomfortably
close to the limits on $f_{\phi}$ from gradient energy density.
In a forthcoming publication \cite{Andriyetal} we argue that, in
general, $\eta$ should be replaced by the quantity
$\tilde{\eta}=\eta+\omega$, and that $\tilde{\eta}$ may be small
even if $\eta$ is not.  The term $\omega$ arises from the terms
of the effective non-renormalizable Lagrangian of the form
$\lambda_{n} m_{\psi} \phi^{2n} \bar{\psi}\psi$ or $\sigma_{n}
m_{\chi}^2 \phi^{2n} \chi^2$ that couple the scalar field $\phi$
to the energy density in other matter fields.  We note in
passing that for those theories that do not lead to `fifth
forces' which violate the weak-equivalence-principle (WEP)
$\tilde{\eta}=0$ precisely.

We thus have a constraint $[\lambda(dg/d\phi)]^{-1} \gtrsim 10^{10}$~GeV to the
scalar-photon coupling, and a constraint $(2\lambda/0.01) \times 10^{-30}$~eV $
\lesssim m_\phi\lesssim
10^{-28}$~eV to the scalar-particle mass.  We note that these
relations may seem hard to reconcile, as contributions to the
scalar-particle propagator from divergent loop diagrams
containing photons should generically be large.  Likewise, if
$\phi$ is a Goldstone mode for some spontaneously broken global
symmetry, it may be unusual from the point of view of
Planck-scale physics  \cite{KamMar92}.  It was also pointed out in
Ref.~\cite{BanDinDou02} that variations of a light scalar should give rise to
large variations in vacuum energy density.  On the other
hand, similar problems arise in models for inflation and for
scalar-field models for dark energy and dark matter.  Here, we
take the view that our $\phi$-photon coupling is simply a
low-energy effective Lagrangian, and anticipate that the light
mass can be protected, e.g., by a mechanism such as that
proposed in Ref. \cite{Arketal03} where the scale of an inflaton
potential is fixed by the size of an extra dimension.  If so
then one might assume that whatever mechanism alleviates the
cosmological-constant problem may also alleviate problems
associated with vacuum-energy gradients.

\section{Conclusion}
\label{sec:Conclusion}

In this paper we have studied the effects on the CMB of spatial fluctuations in
the fine-structure parameter between causally disconnected regions of the
universe at the time of recombination.  Although we have focused on the
particular case of fluctuations in the fine-structure parameter, the formalism
we have
presented may be applied {\it mutatis mutandis} to other modifications of
recombination physics that do not alter the evolution of the dominant density
perturbations.
As discussed above, such fluctuations will alter the predicted CMB power
spectra, introduce a B-mode polarization signal, and introduce temperature and
polarization trispectra and perhaps bispectra.  We stress here that these
results are not dependent on the particular model for $\alpha$ variation
discussed in Section~\ref{sec:ParticleTheory}.

{}From the point of view of effective field theory, variations of the
fine-structure parameter can be phrased in terms of a scalar-photon interaction
Lagrangian
${\cal L}_{\phi\gamma\gamma}$, the parameters of which can be
chosen to be consistent with current experimental limits.
Ultimately, if such
light, cosmologically interesting, scalars do exist in nature their mass must
be protected by some yet unknown mechanism.

\begin{acknowledgments}
We thank A. Cooray and M. Wise for useful discussions, and A. Cooray and M. Kesden for providing the results of their weak-lensing kurtosis calculations for comparison.  KS acknowledges the support of a Canadian NSERC Postgraduate Scholarship.  This work was
supported in part by NASA NAG5-9821, DoE DE-FG03-92-ER40701
and DE-FG03-02ER41215, and NSF PHY00-71856.
\end{acknowledgments}

\end{document}